\documentclass[pdflatex,sn-mathphys-num]{sn-jnl}

\usepackage{graphicx}%
\usepackage{multirow}%
\usepackage{amsmath,amssymb,amsfonts}%
\usepackage{amsthm}%
\usepackage{mathrsfs}%
\usepackage[title]{appendix}%
\usepackage{xcolor}%
\usepackage{textcomp}%
\usepackage{manyfoot}%
\usepackage{booktabs}%
\usepackage{algorithm}%
\usepackage{algorithmicx}%
\usepackage{algpseudocode}%
\usepackage{listings}%
\usepackage{mathtools}
\usepackage[T1]{fontenc}
\usepackage{relsize}
\usepackage{siunitx}
\usepackage{braket}

\newcommand{\Gside}{G_{\mathrm{side}}}

\newcommand{\Grad}{G_{\mathrm{radial}}}
\theoremstyle{thmstyleone}%
\theoremstyle{thmstyletwo}%

\theoremstyle{thmstylethree}%

\renewcommand{\thesection}{Supplementary Note \arabic{section}}
\usepackage{caption}  

\begin{document}

\title[Article Title]{Surface-Code Hardware Hamiltonian}

\author[1]{\fnm{Xuexin} \sur{Xu}}

\author[1]{\fnm{Kuljeet} \sur{Kaur}}

\author[2]{\fnm{Chlo\'e} \sur{Vignes}}

\author[3]{\fnm{John M.} \sur{Martinis}}

\author*[1]{\fnm{Mohammad H.} \sur{Ansari}}\email{m.ansari@fz-juelich.de}

\affil[1]{\orgdiv{PGI-2}, \orgname{Forschungszentrum J\"ulich}, \orgaddress{\city{J\"ulich}, \postcode{52425}, \country{Germany}}}

\affil[2]{\orgdiv{Department of Electrical Engineering and Computer Science}, \orgname{Massachusetts Institute of Technology}, \orgaddress{\city{ Cambridge, Massachusetts}, \postcode{02139}, \country{USA}}}
\affil[3]{\orgname{Qolab}, \orgaddress{ \city{Madison, Wisconsin}, \postcode{53706}, \country{USA}}}


\abstract{
We present a scalable framework for accurately modeling many-body interactions in surface-code quantum processor units (QPUs). Combining a concise diagrammatic formalism with high-precision numerical methods, our approach efficiently evaluates high-order, long-range Pauli string couplings and maps complete chip layouts onto exact effective Hamiltonians. Applying this method to surface-code architectures, such as Google's \emph{Sycamore} lattice, we identify three distinct operational regimes: computationally stable, error-dominated, and hierarchy-inverted. Our analysis reveals that even modest increases in residual qubit–qubit crosstalk can invert the interaction hierarchy, driving the system from a computationally favorable phase into a topologically ordered regime. This framework thus serves as a powerful guide for optimizing next-generation high-fidelity surface-code hardware and provides a pathway to investigate emergent quantum many-body phenomena.}

\maketitle

\section*{Introduction}
\label{sec:intro}

Quantum computing heralds advances in cryptography~\cite{shor1997polynomial-time}, molecular simulation~\cite{aspuru2005simulated}, and combinatorial optimization~\cite{farhi2001quantum}. Yet, this potential is constrained by the fragility of current hardware: qubits---the fundamental carriers of quantum information---are susceptible to rapid decoherence, control imperfections, and parasitic interactions such as stray couplings and crosstalk, all of which degrade both gate and idle fidelities~\cite{mohseni2024,tuokkola2024methods,Benevides_2024,ku2020suppression}.

Quantum error correction~\cite{gottesman1997,kitaev1997quantum,PhysRevLett.129.030501} and error mitigation strategies~\cite{KandalaQEM,scienceadverrormitigation,pazem2025error} offer principled approaches to suppress such errors. However, these protections come at a steep cost: a single logical qubit may require hundreds to thousands of physical qubits, leading to deeper circuits and placing substantial demands on fabrication yield, control electronics, and cryogenic infrastructure. The central engineering challenge is therefore to reduce physical error rates while containing overhead, so that the full promise of quantum advantage becomes practically realizable~\cite{Preskill2018quantumcomputingin,mohseni2024}.

As quantum processing units (QPUs) scale, tighter fabrication tolerances and inevitable defects underscore the importance of predictive modeling. Pre-mask simulations have become indispensable, enabling the identification and mitigation of parasitic interactions---such as stray couplings---prior to lithographic fabrication. This rising demand converges toward a central question: \emph{Can we simulate QPUs with pristine and predictive accuracy?} Addressing this question calls for a brief reflection on the remarkable history of theoretical advances that have enabled qubit architectures and quantum processors to reach their current level of technological maturity~\cite{krantz19a-quantum,ciani2023lecture}.

Early circuit-QED models, restricted to pairwise couplings, proved valuable in guiding experiments with gate fidelities worse than $99\%$. However, as gate errors approach the $0.1\%$ level, such simplified Hamiltonians have become inadequate: they provide only a coarse approximation of the underlying physics and fail to meet the stringent accuracy now demanded by quantum hardware~\cite{devoret2013superconducting,koch2007charge-insensitive}. Subsequent exact, non-perturbative analyses have shown that even ideal two-body models fall short of current fidelity benchmarks~\cite{ansari19superconducting}. 

To overcome these limitations, lattice-Hamiltonian approaches have emerged, incorporating full three-body interactions within three-qubit unit cells~\cite{xu24lattice}. These methods have uncovered novel many-body phases---including hierarchy-inverted and near-chaotic regimes---that coexist with or border the computational phase~\cite{berke2022transmon,xu24lattice}.

Among quantum error-correcting architectures, the \emph{surface code} remains the leading candidate for fault-tolerant quantum computing, owing to its high threshold (approximately $1\%$ per gate), modest stabilizer overhead, and reliance solely on nearest-neighbor couplings~\cite{acharya23suppressing,acharya25quantum}. Logical information is encoded nonlocally across a two-dimensional lattice of data and ancilla qubits, allowing local faults to be detected and corrected before they propagate. The code’s patch-based modularity enables scalable tiling of logical qubits, offering a clear blueprint for constructing universal, fault-tolerant processors. Recent experimental milestones---most notably Google’s \emph{Willow}---have expanded surface code lattices while concurrently lowering logical error rates, illustrating how this architecture can transform today’s noisy devices into reliable platforms operating well below the fault-tolerance threshold~\cite{acharya25quantum}.

We employ the method of \emph{diagrammatic-perturbation} that reconstructs the full Hamiltonian of a quantum-processing unit (QPU) residing in the Hilbert space of \emph{all-to-all} coupled qubits, irrespective of whether the couplings are deliberately engineered or merely parasitic. Valid at arbitrary perturbative order and for interactions of any rank, the formalism yields closed-form coefficients that flow directly into a high-precision layout-to-Hamiltonian estimator. Building on our earlier three-qubit study~\cite{xu24lattice}, we extend the approach to the five-qubit surface-code tile to resolve Pauli-string couplings with sub-kHz accuracy. The resulting Hamiltonian lets us toggle individual qubit–qubit interactions on and off and quantify many-body effects on gate activation and deactivation. 

We next apply the framework to Google’s \emph{Sycamore} lattice, whose publicly available parameters are incomplete, and uncover three distinct computational regimes: a benign phase dominated by pairwise $ZZ$ couplings~\cite{xu2021zz-freedom,xu2020experimental}, a parasitic phase in which  three-body $ZZZ$ terms prevail~\cite{xu24lattice,xu2023parasitic-free}, and a chaotic domain characterized by weak $ZZ$ yet strong $ZZZ$ interactions~\cite{xu24lattice,berke2022transmon,fischer2024dynamical}. Even infinitesimal drifts in residual qubit–qubit couplings can trigger phase transitions, invert the coupling hierarchy, and precipitate sharp drops in two-qubit-gate fidelity. To benchmark these effects, we analyze the native {iSWAP} gates of the processor, tracking how residual errors and many-body crosstalk propagate throughout the QPU. Leveraging the predictive power of our Hamiltonian, we virtually optimize residual couplings, gate scheduling, and control pulses--well before fabrication--narrowing the gap between theoretical error-correction protocols and experimental implementations. Finally, the framework provides universal diagnostics that detect and suppress many-body error channels early in the design cycle, thereby accelerating the emergence of quantum processors capable of high-fidelity simulation, secure communication, and large-scale optimization.

\section*{Results}\label{subsec:cell}
\subsection*{Hamiltonian of QPU}

\begin{figure*}[ht]
    \centering
    \includegraphics[width=0.95\linewidth]{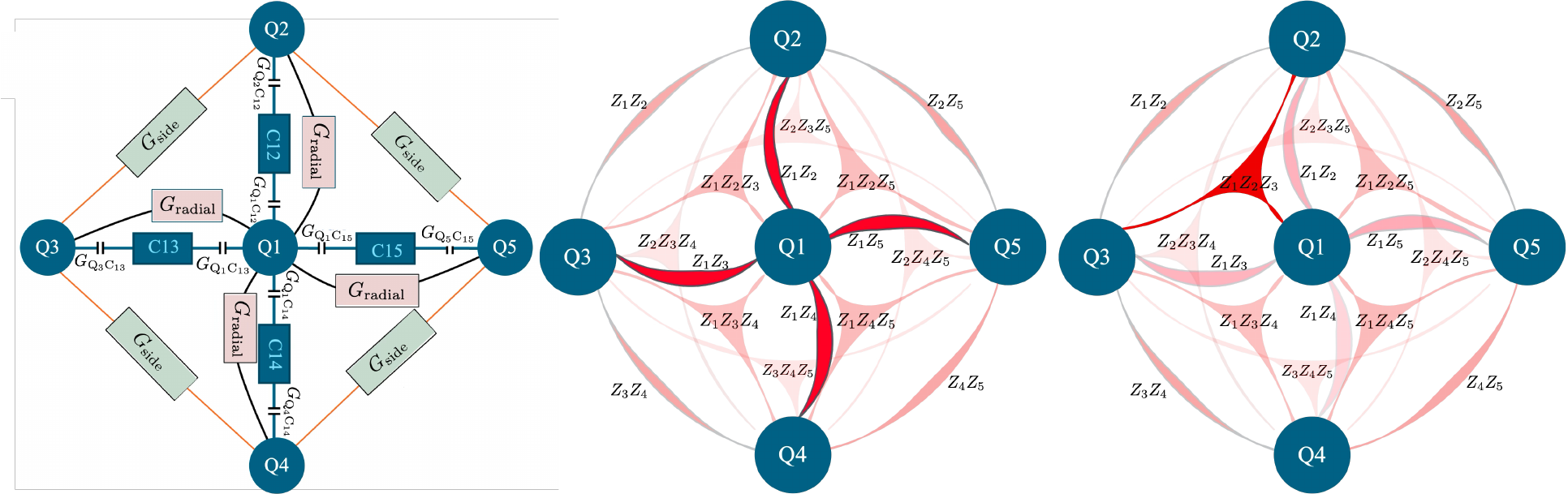}\put(-355,110){\textbf{(a)}}\put(-230,110){\textbf{(b)}}\put(-115,110){\textbf{(c)}}
	\caption{Schematic of a standard five-qubit surface-code unit cell in a diamond lattice. (a) A central measure qubit $Q_1$ couples four surrounding data qubits $Q_{2}$–$Q_{5}$ via dedicated tunable couplers $C_{1p}$ ($p=2,3,4,5$). Direct capacitive couplings include a universal central-to-data links $\Grad$ and data-to-data couplings $\Gside$. (b) A representative perturbative hierarchy where side-to-central qubit $ZZ$ interactions dominate over side   couplings as well as three-body $ZZZ$ interaction. (c) $ZZZ$ superiority involving the central qubit dominates over two-qubit interactions.}
    \label{fig:unit_cell}
\end{figure*}

In the dispersive limit, we use diagrammatic perturbation theory to derive closed-form estimates of the interaction strengths, supplying priors for subsequent high-precision numerical evaluation \cite{cirqubit}, details are given in \textbf{Method}. For the unit cell shown in Fig.~\ref{fig:unit_cell}(a), the Hamiltonian is governed by four \emph{radial} $ZZ$ couplings and six \emph{lateral} $ZZ$ couplings that link nearest- and next-nearest-neighbor qubits.  Beyond these pairwise terms, three-body $ZZZ$ interactions arise, either involving the central qubit together with two outer qubits or exclusively among the outer qubits, i.e.
\begin{align}
H_{\text{cell}}  &= \sum_{i=1}^{5} H_{Q_i}
            + \sum_{p}\alpha_{p1}\,Z_pZ_1    + \!\!\!\sum_{\langle p,q\rangle; \;\langle\!\langle p,q\rangle\!\rangle}\!\!\!\alpha_{pq}\,Z_pZ_q                              \notag\\
          &\quad 
            + \!\!\!\sum_{\langle p,q\rangle;\;\langle\!\langle p,q\rangle\!\rangle}\!\!\!\alpha_{pq1}\,Z_pZ_qZ_1
            + \!\!\!\sum_{\;\langle\!\langle p,q,r\rangle\!\rangle}\!\!\!\alpha_{pqr}\,Z_pZ_qZ_r,
\label{eq:latticeHam}
\end{align}
and throughout this work we write $\langle\cdot\rangle$ for nearest-neighbor (NN) qubit pairs and $\langle\!\langle\cdot\rangle\!\rangle$ for next-nearest neighbors (NNN).  Our definition of neighborhood is deliberately topological rather than metric: we delete every coupler from the circuit diagram and examine the graph formed solely by qubit vertices.  Two qubits are deemed NNs if, in this reduced lattice, no additional qubit lies between them---irrespective of the microscopic interaction pathway or their Euclidean separation on chip.  Consequently, even in the absence of a direct coupler, qubits that occupy adjacent sites of the pruned lattice are classified as NNs.  This convention divorces the notion of neighborhood from hardware-specific wiring and instead focuses on the connectivity that ultimately dictates error propagation and code performance.

Coupler-qubit interactions $G_{QC}$ are the strongest coupling in the hierarchy, followed by NN couplings $G_{\langle QQ\rangle}$, and finally NNN, $G_{\langle\!\langle QQ\rangle\!\rangle}$, establishing the hierarchy
$ G_{QC}\!\gg\!G_{\langle QQ\rangle}\!\gg\!G_{\langle\!\langle QQ\rangle\!\rangle}$. Consequently, two-body \emph{radial} terms ($\alpha_{p1}$) outweigh \emph{lateral} interactions between both closest and more distant NN's; the same ordering carries over to the associated three-body coefficients.  
Adopting this hierarchy and treating all non-participating qubits as idle (\(I\)), we derive explicit expressions for the relevant \(ZZ\) and \(ZZZ\) couplings within the lattice.

Let us now determine the two- and three-body Pauli strengths explicitly.  The Pauli interaction of two bodies $Z_iZ_j$ between ${Q_i}$, ${Q_j}$ up to the third perturbative order is $\alpha_{ij} = \alpha_{ij}^{(2)} +  \alpha_{ij}^{(3)} + O(4)$ with the following details for each according to the perturbative expansion:
\begin{align}
 \frac{\alpha_{ij}^{(2)}}{2} \equiv & \frac{{(J_{i\overline{j}}})^2}{\Delta_{i\overline{j}}} - \frac{{(J_{\overline{i}j}})^2}{\Delta_{\overline{i}j}} \\ 
\frac{\alpha_{ij}^{(3)}}{4} \equiv &\sum_k\! \frac{J_{ij} J_{ik} J_{jk}}{\Delta_{ik} \Delta_{jk}}\!+\!\frac{J_{ij} J_{i\overline{k}} J_{j\overline{k}}}{\Delta_{i\overline{k}} \Delta_{j\overline{k}}}\!  -\! \frac{J_{\overline{i}j} J_{\overline{i}k} J_{jk}}{\Delta_{\overline{i}k} \Delta_{jk}}\! -\! \frac{J_{i\overline{j}} J_{ik} J_{\overline{j}k}}{\Delta_{ik} \Delta_{\overline{j}k}}\nonumber \\ &
+2\left( \frac{J_{\overline{i}\overline{j}} J_{\overline{i}k} J_{\overline{j}k}}{\Delta_{\overline{i}k} \Delta_{\overline{j}k}}+ \frac{J_{\overline{i}\overline{k}} J_{\overline{i}j} J_{\overline{k}j}}{\Delta_{\overline{i}j} \Delta_{\overline{k}j}} 
+ \frac{J_{\overline{j}\overline{k}} J_{\overline{j}i} J_{\overline{k}{i}}}{\Delta_{\overline{j}{i}} \Delta_{\overline{k}{i}}} \right)
 \label{eq:zz2}
\end{align}
with $\Delta_{ij} \equiv f^{1\to 0}_i - f^{1\to 0}_j$  and $\Delta_{i\overline{j}} \equiv f^{1\to 0}_i - f^{2\to 1}_j$. The third-order correction involves a summation over all intermediate qubits $k \ne i, j$.

The three-body interaction $Z_iZ_jZ_k$ between ${Q_i}$, ${Q_j}$, and ${Q_k}$  can be obtained by considering the reference levels $|\cdots 1_i, 1_j, 1_k, \cdots\rangle $ with $\cdots$ representing other qubit states, each of which could be $|0\rangle$ or $|1\rangle$. Considering all such states, one can determine that the lowest perturbative order for $ZZZ$ is $J^3/\Delta^2$. In the following we find it in its first order, that is, $\alpha_{ijk}=\alpha_{ijk}^{(3)}+O(4)$ and this in detail is the following: 
\begin{align}
\frac{\alpha_{ijk}^{(3)}}{8} = &-\left( \frac{J_{ij} J_{i\overline{k}} J_{j\overline{k}}}{\Delta_{i\overline{k}} \Delta_{j\overline{k}}}+ \frac{J_{ik} J_{i\overline{j}} J_{k\overline{j}}}{\Delta_{i\overline{j}} \Delta_{k\overline{j}}}+ \frac{J_{jk} J_{j\overline{i}} J_{k\overline{i}}}{\Delta_{j\overline{i}} \Delta_{k\overline{i}}}\right. \nonumber \\ 
&\left. \quad \quad + \frac{J_{\overline{i}\overline{j}} J_{\overline{i}k} J_{\overline{j}k}}{\Delta_{\overline{i}k} \Delta_{\overline{j}k}}+ \frac{J_{\overline{i}\overline{k}} J_{\overline{i}j} J_{\overline{k}j}}{\Delta_{\overline{i}j} \Delta_{\overline{k}j}}   + \frac{J_{\overline{j}\overline{k}} J_{\overline{j}i} J_{\overline{k}{i}}}{\Delta_{\overline{j}{i}} \Delta_{\overline{k}{i}}} \right)
\label{eq.ZZZ}
\end{align}

Exact numerical simulations~\cite{cirqubit} reveal that when the detuning approaches resonance, $\Delta\!\approx\!0$, the usual hierarchy of pairwise and three-body couplings can invert, ushering the processor into a regime dubbed \emph{$ZZZ$ superiority}~\cite{xu24lattice}. Crucially, this inverted phase is \emph{not} confined to the resonance line; it occupies a finite swath of parameter space, making it crucial to differentiate the true quantum-computing phase--where the conventional ordering $|ZZ|\!\gg\!|ZZZ|$ prevails--from neighboring exotic phases that, while fascinating for many-body physics, severely degrade gate fidelity. Figure~\ref{fig:unit_cell}(b) sketches the standard perturbative hierarchy: Strong radial $ZZ$ couplings bind each side qubit to the central qubit, lateral $ZZ$ links among side qubits are weaker, and three-body terms involving the center dominate those acting solely on side qubits. Figure~\ref{fig:unit_cell}(c) contrasts this with the $ZZZ$-superiority regime, where the many-body ordering collapses and three-qubit interactions overshadow their pairwise counterparts.

As quantum processors grow in size, the proliferation of many-body interactions---and the accompanying breakdown of many-body localization---forces higher-order effects into the spotlight.  This is especially true for devices such as Google’s \emph{Sycamore}, where aggressive tuning of couplings introduces pronounced disorder and can generate unexpected hierarchies among many-body terms.

\subsection*{Toggling Coupler $J$ ON/OFF}
Experimental data show that pushing residual Pauli couplings below the {50} kilohertz threshold essentially eliminates their impact on gate fidelity limit required for fault tolerance.  A back‐of‐the‐envelope estimate supports this: a stray \(ZZ\) term adds an unwanted phase,
\(\ket{11}\!\mapsto\!e^{i\alpha_{ZZ}t}\ket{11}\), so during a {50} nanosecond gate length the state fidelity falls by roughly \(5\times10^{-4}\).  Because the coupling remains active during idle periods, the same error is always-on and builds up coherently, and the accumulated infidelity must be countered by mitigation protocols, such as those in Ref.~\cite{pazem2025error} and in the references therein. 

Although recent hardware advances have driven residual couplings ever lower, a complementary line of research seeks to \emph{exploit} regimes where the usual interaction hierarchy breaks down in order to simulate quantum chaos. Credible tests of this idea demand Hamiltonians that capture realistic device parameters and encompass many‐body terms rather than merely pairwise couplings. ~\cite{berke2022transmon,fischer2024dynamical}

\begin{figure}[t]
    \centering
    \includegraphics[width=0.45\linewidth]{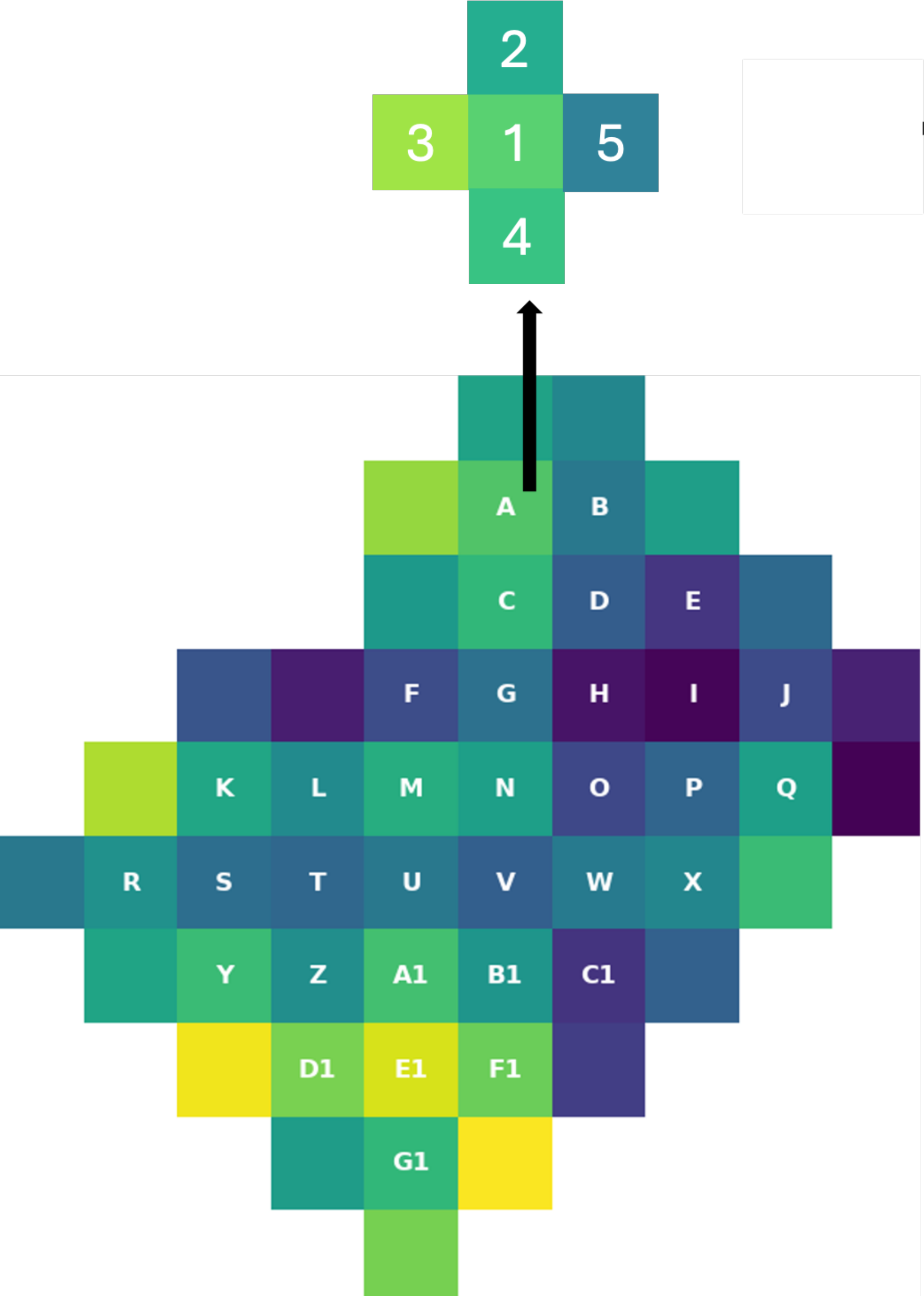}
\caption{Labeling convention of unit cells on the Sycamore processor~\cite{arute19quantum}. Each unit cell consists of a central qubit, labeled as $Q_1$, surrounded by four peripheral qubits, labeled $Q_2$–$Q_5$ in a counterclockwise arrangement. The colors indicate typical distribution of qubit frequency over the Sycamore processor.}  
    \label{fig:sycamore}
\end{figure}

We model Google’s \textit{Sycamore} processor using the circuit parameters reported in Ref.~\cite{arute19quantum}. Figure~\ref{fig:sycamore} depicts the qubit spectrum: qubits rendered in the same color share a common frequency, whereas contrasting colors mark detunings large enough to justify a perturbative treatment of residual couplings. The 53-qubit lattice partitions naturally into thirty-three overlapping surface-code unit cells, each a five-qubit diamond as shown in Fig.~\ref{fig:unit_cell}; the central qubits of these diamonds are labelled A–G1 for later reference. From this layout we construct a Hamiltonian that retains both pairwise \(ZZ\) and three-body \(ZZZ\) interactions, providing a realistic platform for exploring chaotic dynamics under experimentally relevant conditions.

We first park every qubit and coupler at its reported idle frequency, canceling all $J$'s.   To isolate one tunable link, we sweep only the $C_{13}$ coupler between $Q_1$ and $Q_3$, while \emph{keeping qubit frequencies fixed}.   This protocol differs from an iSWAP gate, where qubit detuning is co-modulated with the coupler bias; here we study the bare act of turning $J_{13}$ on and off.

At each bias point, we recompute the full lattice Hamiltonian, tracking how the $C_{13}$ sweep perturbs every two- and three-body coefficients, both inside and outside the computational subspace.  
If non-gate $ZZ$ or $ZZZ$ terms reappear, the remaining couplers are rebiased to restore their cancelation.  
The resulting high-dimensional optimization is impractical to perform experimentally in real time but is readily handled in simulation, enabling rapid exploration of error-minimizing bias schedules.

\begin{figure}[ht]
    \centering
    \includegraphics[width=0.95\linewidth]{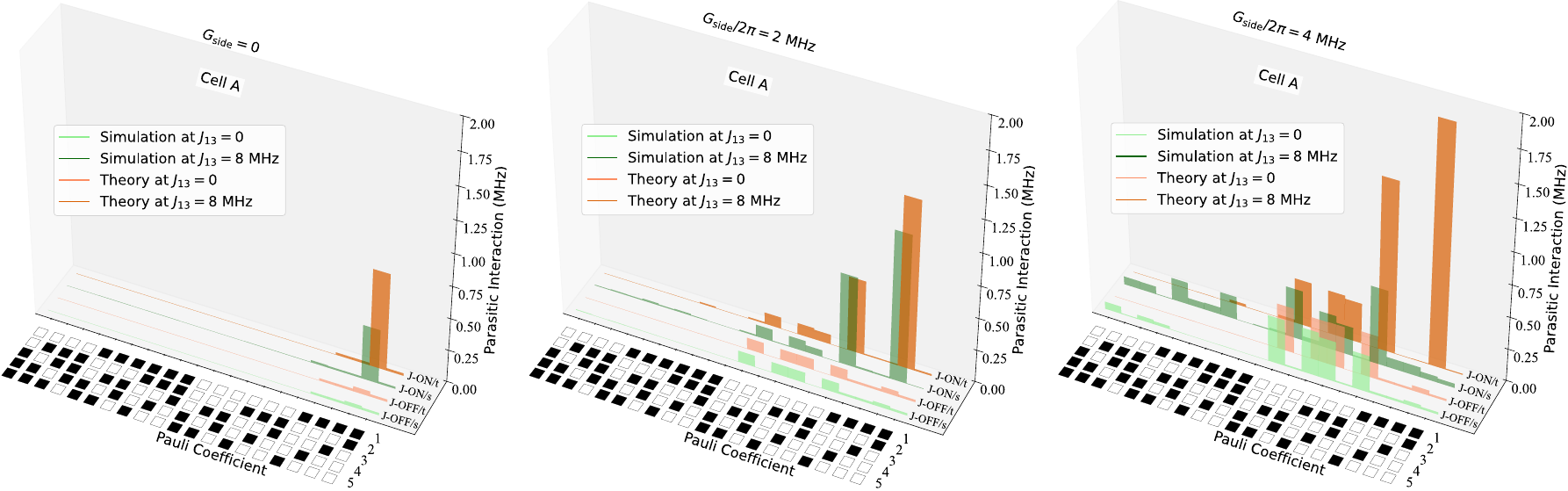}\put(-355,110){\textbf{(a)}}\put(-237,110){\textbf{(b)}}\put(-117,110){\textbf{(c)}}

    \caption{Hamiltonian tomography of parasitic interactions for both the effectively off and on states of unit cell A, with $J_{13} = 0$ (light color) and $J_{13}/2\pi = 8$ MHz (deep color), respectively. The x-axis represents 20 different parasitic interactions, where two- and three-qubit Pauli-$Z$ terms are filled in black, while Pauli-$I$ terms remain unfilled. Such parasitic interactions are calculated under three scenarios (a) $G_{\text{side}}=0$, (b) $G_{\text{side}}/2\pi=2$ MHz and (c) $G_{\text{side}}/2\pi=4$ MHz with results from perturbation theory shown in orange and those from numerical simulations shown in green.} 
    \label{fig:chaos4}
\end{figure}

To explore interaction-driven disorder, we introduce a universal side   coupling $G_{\text{side}}$.
Throughout this work, each qubit remains at its experimentally determined idle frequency, while every coupler is biased into the hard-decoupling ($J$-OFF) regime described in Ref.~\cite{xu24lattice}. When we later discuss gate operations, we also consider a soft-decoupling setting in which $ZZ \simeq ZZZ \simeq 0$ and the couplers operate in the $J$-$\widetilde{\text{OFF}}$ configuration. Direct radial couplings $G_{Q_1Q_i}$, collectively denoted $\Grad$, merely shift the zero of the composite interaction $J_{Q_1Q_i}$ once the tunable path through $C_{1i}$ is taken into account and therefore do not require further treatment.

Figure~\ref{fig:chaos4} displays the amplitudes of all five-qubit Pauli strings $\hat{O}_1\hat{O}_2\hat{O}_3\hat{O}_4\hat{O}_5$, with each single-qubit operator $\hat{O}\in{\hat{I},\hat{Z}}$ and at most three instances of $\hat{Z}$. Filled squares mark qubits acted on by $\hat{Z}$, whereas open squares correspond to identities. Every simulation fixes the direct couplings at $G_{Q_iC_{i1}}/2\pi = 100~\text{MHz}$ and employs a uniform side coupling $\Gside$. The values of $G_{Q_iC_{i1}}/2\pi$ and $\Gside$ are based on rough estimates reported in Ref.~\cite{PhysRevLett.125.120504}.

We begin with all $J$ couplings turned off--including the target coupling $J_{13}$, namely $J$-OFF.  Adiabatically we ramp the control parameter of coupler $C_{13}$ until the desired $J_{13}$-on configuration is reached. After each incremental step we recompute the full lattice Hamiltonian and retune the remaining couplers so that every $J$ except $J_{13}$ stays off. This protocol enables us to quantify how activating a single coupler while holding all others off influences stray interactions. Numerical results (green bars) agree with third-order perturbation theory (orange) for small $J_{13}$, but as $J_{13}$ increases the analytical estimate progressively underestimates parasitic errors. Moreover, as $\Gside$ is increased, strong three-body $ZZZ$ couplings emerge and can dominate the Hamiltonian when near-resonant settings are used to explore quantum chaos~\cite{berke2022transmon}.

\begin{figure*}[ht]
    \centering
     \includegraphics[width=1\linewidth]{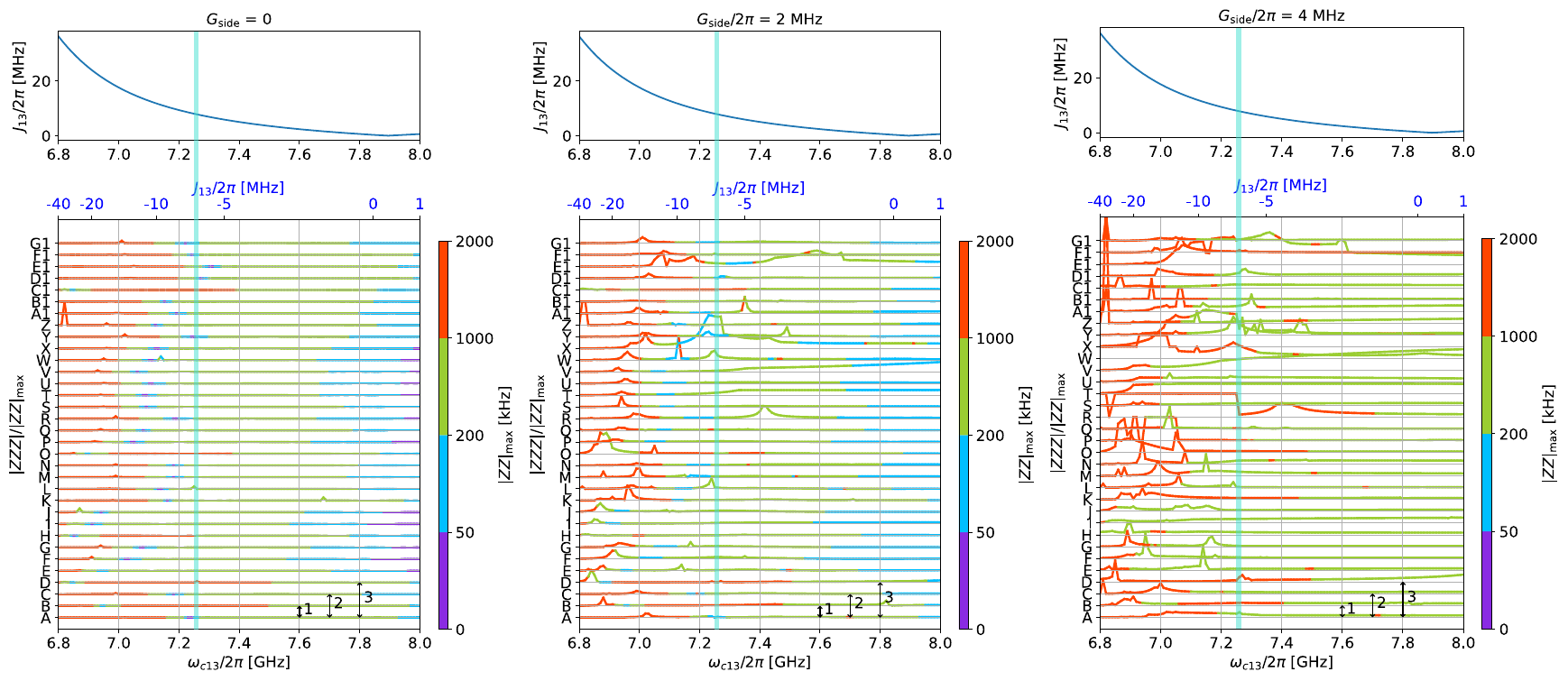}\put(-375,160){\textbf{(a)}}\put(-255,160){\textbf{(b)}}\put(-130,160){\textbf{(c)}}

\caption{Processor Error Tomography (PET) of the parasitic interactions across each cell calculated using CirQubit. \textbf{Top:} Effective $J_{13}$ coupling as a function of coupler $C_{13}$ frequency. \textbf{Bottom:} Pauli channels for all unit cells. Curves arranged along the rows indicate the ratio $|ZZZ|_{\rm max}/|ZZ|_{\rm max}$ for each unit cell. Color codes represent the magnitude of the maximum $ZZ$ interaction. A ratio greater than one indicates that the three-body $ZZZ$ interaction dominates over the two-body $ZZ$ interaction at different side couplings (a) $\Gside=0$, (b) $\Gside/2\pi=2$ MHz and (c) $\Gside/2\pi=4$ MHz. } 
\label{fig:earthquake4}
\end{figure*}

Using full-device simulation by Ref.~\cite{cirqubit} we find that chaotic behavior emerges once the side–to–side coupling exceeds $G_{\text{side}}/2\pi\gtrsim4\;\text{MHz}$ (Fig.~\ref{fig:chaos4}).  We evaluated parasitic interactions for two distinct states: the effectively OFF state ($J_{13}=0$, light color) and the ON state ($J_{13}/2\pi=8$ MHz, deep color) for three different $G_{\text{side}}/2\pi=0,2,4$ MHz, separately. Perturbative calculations (orange) are compared to numerical simulations (green).

Both numerical simulations and perturbative analysis demonstrate that, with every coupler OFF, parasitic interactions essentially disappear--apart from nearest-neighbor two-body strings $Z_jZ_{j+1}$ for $j \in {2,3,4,5}$. However, increasing solely the $J_{13}$ coupling introduces marked disorder. Although most couplings remain modest, several three-body terms unexpectedly rise to the few-hundred-kilohertz level.

In quantum computing each unit cell operates as an independent module, so we must first establish the hierarchy of interactions inside every cell.
Doing so allows us to pinpoint cells that display anomalous numerical behavior during computation.
We therefore condense the interaction data calculated using CirQubit for each cell and stack the resulting images into a single composite graph, which we call {processor error tomography} (PET) and present in Fig.~\ref{fig:earthquake4}.
In a PET, each row represents one unit cell and encodes the ratio $|ZZZ|_{\text{max}}/|ZZ|_{\text{max}}$, while a color map simultaneously reports the largest two-body $ZZ$ coupling.
A ratio greater than unity signals that three-body interactions dominate.
Consequently, a PET offers a lattice-wide view of stray couplings across multiple Pauli channels, with the vertically stacked rows corresponding to successive five-qubit tiles.
Figure~\ref{fig:earthquake4} also aligns the coupler frequency $\omega_{C_{13}}$ along the upper horizontal axis with the corresponding strength $J_{13}/2\pi$ on the lower axis.
As we sweep the frequency, the coupling hierarchy evolves: smooth, blue traces delineate parameter windows where $|ZZZ|_{\text{max}}<|ZZ|_{\text{max}}$, and the blue hue marks low absolute $ZZ$ coupling within the cell.
At $J_{13}/2\pi = 8~\text{MHz}$ (green vertical line) and $\Gside/2\pi = 4~\text{MHz}$, however, pronounced disorder appears, most prominently in unit cell \textbf{R}; additional details are given in the supplementary.
Although a PET cannot reveal which specific interaction is at fault, it nevertheless provides a succinct survey of the parasitic budget of every tile at any frequency and thus helps identify operating regimes in which all cells attain higher fidelity.

\subsection*{iSWAP Gate Performance}

In this section, we benchmark the iSWAP gate realized on the Sycamore superconducting processor, emphasizing how the achieved gate fidelity depends on spectator-qubit (side-qubit) coupling as well as on higher-order parasitic interactions inherent to the hardware.  
For this purpose, we treat the calibrated iSWAP pulse sequence as the sole intentional two-qubit operation and--purely for conceptual clarity--do not analyze the automatically generated $ZZ$ correction separately, even though compensation is routinely employed in fermionic-simulation protocols and during concurrent execution of fSim~\cite{jiang2024concurrent,barends2015digital}. 

\subsubsection*{Benchmark Protocol}

To implement the iSWAP gate between $Q_{1}$ and $Q_{3}$, we tune $Q_{3}$ until its transition frequency equals that of $Q_{1}$, $\tilde{\omega}_{1}=\tilde{\omega}_{3}$, while disabling all other tunable couplings of $Q_{1}$.  
The detuning that $Q_{1}$ would otherwise experience is expected to differ markedly, a scenario that we do not investigate here.

\begin{figure*}[ht]
    \centering
     \includegraphics[width=0.85\textwidth]{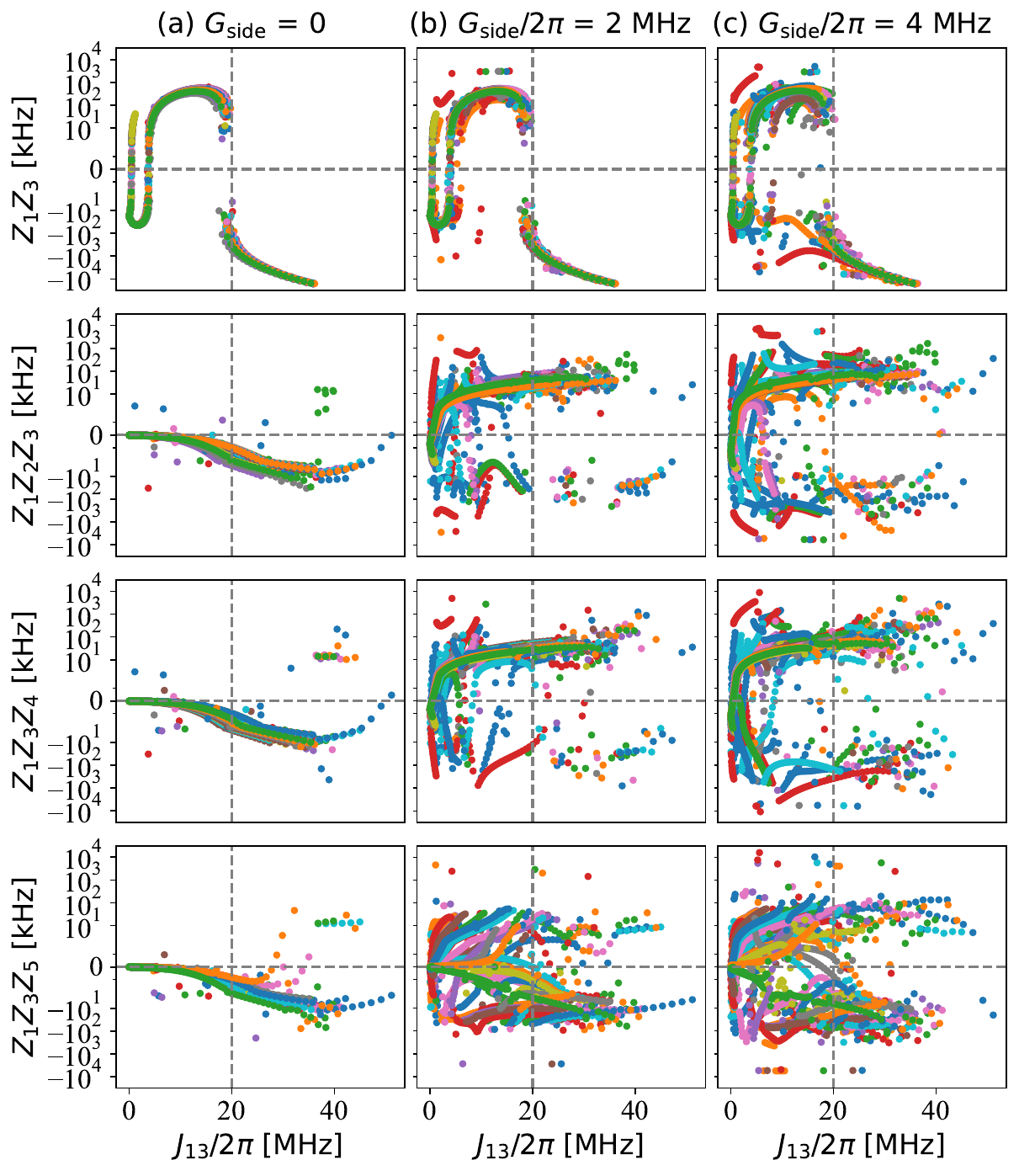}

\caption{The overlay of $Z_1Z_3$, $Z_1Z_2Z_3$, $Z_1Z_3Z_4$, and $_1Z_3Z_5$ parasitic interactions for all 33 unit cells used in Sycamore processor, plotted against effective iSWAP gate strength $J_{13}$ for On and OFF gate between $Q_1$ and $Q_3$, evaluated using CirQubit}. The side qubit-qubit coupling $\Gside/2\pi$ is zero in (a), 2 MHz in (b), and 4 MHz in (c).
\label{fig:pzz0}
\end{figure*}

The qubit-idle-frequency data we presented in the previous section assumed that qubits remain parked at their frequency.  However, to accommodate the new resonance condition, we recomputed the engineered exchange $J_{13}$ together with the accompanying parasitic interactions.  

Figure~\ref{fig:pzz0} (a-c) shows the updated stray coupling landscape for the resonant configuration at three values of $\Gside /2\pi= 0,\,2,\,4~\text{MHz}$, plotted against the effective exchange coupling between two qubits, $J_{13}$, based on the calculation using CirQubit.  We overlay stray-coupling curves from \emph{all} unit cells to facilitate direct cell-to-cell comparison.  
The resulting envelopes delineate operating windows in which the gate is unambiguously \textsc{on} (maximal exchange) or \textsc{off} (suppressed exchange), providing practical guidance for pulse scheduling in larger-scale circuits.  

Subsequently, we fine-tune each unit cell to maximize fidelity, obtaining an optimal exchange strength in the range $J_{13}/2\pi \approx 18\text{–}24~\text{MHz}$; the precise value depends on the particular cell, the numerical optimization routine, and the fixed side coupling amplitude $\Gside$.

We optimize the exchange coupling $J_{13}$ in each unit cell to maximize the fidelity of iSWAP under three progressively more realistic noise models: (i) intrinsic decoherence only; (ii)  decoherence plus parasitic two-body $ZZ$ couplings;  (iii) decoherence, two-body $ZZ$ couplings, and three-body $ZZZ$ interactions. For every model, we quote the minimum gate error obtained at the corresponding optimal $J_{13}$ of the noise model (ii). All simulations target the \textsc{Sycamore} architecture, and the resulting fidelity of the $\mathrm{iSWAP}_{Q_1Q_3}$ gate is reported as a function of the side coupling strength $\Gside$.

We probed the sensitivity of the iSWAP to the control envelope using a pulse \emph{flat-top Gaussian}: a short Gaussian
ramp–up, a constant plateau at the target frequency, and a symmetric ramp-down (see more details in Supplementary Fig. S2). The same template is applied to every unit cell, but the exact width is adjusted to the locally optimized exchange coupling $J_{13}$.

\begin{figure*}[ht]
    \centering
    \includegraphics[width=0.7\linewidth]{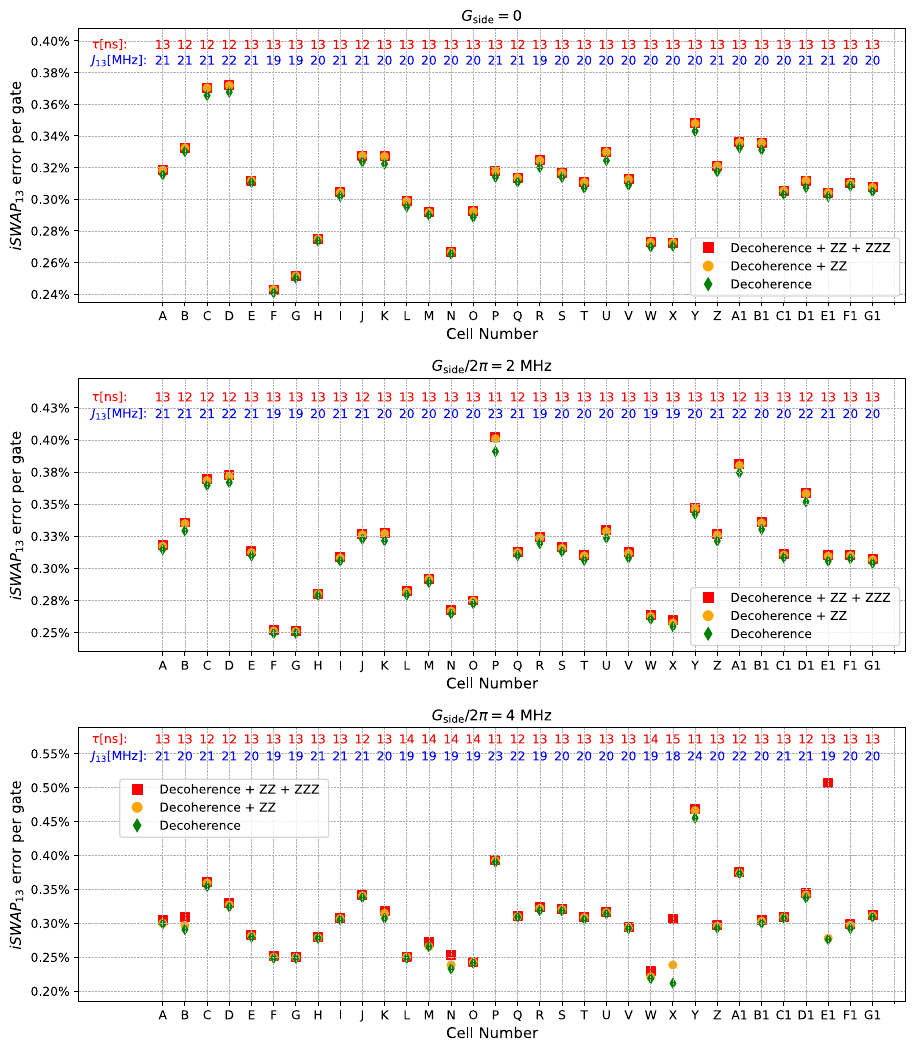}\put(-275,290){\textbf{(a)}}\put(-275,190){\textbf{(b)}}\put(-275,90){\textbf{(c)}}\\

    \caption{Per–cell iSWAP error on \textit{Sycamore} at $\Gside/2\pi=0,2,4$ MHz, separately.  Green diamonds: coherence limit.  
             Orange circles: errors from decoherence + $ZZ$.  
             Red rectangles: decoherence + $ZZ$ + $ZZZ$. }
    \label{fig:iswap_fidelity4}
\end{figure*}

Figure~\ref{fig:iswap_fidelity4}\,(a–c) reports the resulting
gate error for three representative side coupling strengths
$\Gside$.
For each cell we compare  \textcolor{green}{$\blacklozenge$}~decoherence only,  \textcolor{orange}{$\bullet$}~decoherence\,+\,$ZZ$,  \textcolor{red}{$\blacksquare$}~decoherence\,+\,$ZZ$+$ZZZ$.
In all cells, the flat-top Gaussian maintains high fidelity, and the modest spread in pulse length highlights the tolerance of the protocol to local variations in $J_{13}$.

\subsubsection*{Diagnosing Faulty Cells}
\label{subsec:bad_cells}

To pinpoint the origin of outlier performance, we examine the spectrum of parasitic interactions extracted for
every unit cell.  Figure~\ref{fig:pzz0} overlays, for each unit cell, the
magnitudes of the dominant
two-body and three-body  stray interactions evaluated in the locally optimized exchange coupling $J_{13}$.  Two representative
problem cells as shown in Fig.~\ref{fig:iswap_fidelity4}(a-c) are: cells \textbf{X} and \textbf{E1}. These unit cells exhibit the largest discrepancies in their estimated parasitic sensitivity for the case of $\Gside/2\pi=4$ MHz, signaling a pronounced sensitivity to three-body interactions and hence to chaotic dynamics (see more details in Supplementary Fig. S3). Let us look into each one in rather more detail:

\begin{itemize}
\item \textbf{Cell X} ($J_{13}/2\pi \approx 18~\mathrm{MHz}$):
The parasitic couplings of two and three bodies swell into the hundreds of kilohertz range (see more details in Supplementary Fig. S1).
These sizable coherent phase shifts overwhelm the intrinsic decoherence floor, driving the total gate error much greater than the $10^{-3}$ threshold demanded by surface code operation.

\item \textbf{Cell E1} ($J_{13}/2\pi \approx 19~\mathrm{MHz}$):
Although the two-body $ZZ$ channel remains modest ($<50~\mathrm{kHz}$), the three-body interaction $Z_1Z_2Z_3$ alone rises above $300~\mathrm{kHz}$.
This single $ZZZ$ term--even when all pairwise $ZZ$ couplings appear well controlled--provides a crucial lesson: the full fidelity loss seen in Fig.~\ref{fig:iswap_fidelity4} highlights the severe impact that higher-order processes can exert on iSWAP gate performance in the QPU.
\end{itemize}

\medskip

In summary, meticulous two-qubit tuning alone is insufficient: precise
control of \emph{both} $ZZ$ and $ZZZ$ terms is indispensable to realize high-fidelity iSWAP gates across an extended Sycamore-class
device.  Future work will integrate parasite-aware calibration loops
with adaptive pulse shaping, with the aim of pushing the effective three-body
coupling below the $50~\mathrm{kHz}$ threshold that experimentally
marks the onset of fidelity degradation.

\section*{Small is Big!}

The foregoing analysis yields a pivotal insight: once the side-qubit coupling $\Gside$ drifts beyond a modest threshold--about $2~\mathrm{MHz},(2\pi)$--the natural hierarchy $\lvert ZZ\rvert>\lvert ZZZ\rvert$ can invert, even when all two-body stray couplings remain benign. The three-body terms then move to the center stage, and a model limited to pairwise interactions would not allocate an error budget to $ZZ$, thus obscuring the true source of gate infidelity.

Figure~\ref{fig:earthquake4} makes the danger explicit.  With $\Gside=0$, panel (a) places every unit cell in a nearly quiescent regime: both two- and three-body parasitic couplings stay below $100~\mathrm{kHz}$ over a wide span of qubit frequencies (blue and purple regions).  However, increasing $\Gside$ to a modest $2~\mathrm{MHz}$, panel (b) reveals pronounced oscillations in the maximal ratio $|ZZZ/ZZ|$; at selected coupler detuning, the expected ordering $|ZZ|>|ZZZ|$ collapses and the three-body channel overtakes its two-body counterpart.  Even a slight boost in direct side coupling thus amplifies parasitic interactions, such as two-body, three-body, or both, underscoring the delicate balance required for high-fidelity operation.

The error amplification intensifies almost universally when $\Gside$ reaches $4~\mathrm{MHz}$, as illustrated in panel (c), shrinking the safe operating windows for nearly every unit cell and pushing the architecture toward the brink of error correction thresholds.

A similar pattern emerges when the qubit $Q_3$ is tuned to resonance with $Q_1$, as illustrated for individual cells in Fig.~\ref{fig:pzz0}(a–c).  In the absence of side coupling, panel (a) provides broad, well-defined ON and OFF operating windows with negligible parasitic interactions.  In contrast, introducing $\Gside/2\pi = 2$ or $4~\mathrm{MHz}$--panels~(b) and~(c)--compresses these safe windows and makes them acutely sensitive to coupler detuning.  Collectively, these observations suggest that \textbf{ rather than completely eliminating side couplings}, retaining them at small but finite values and explicitly incorporating their effects into simulation can produce more optimal circuit layouts and gate-site selections.

Figure~\ref{fig:iswap_fidelity4} already shows that increasing $\Gside$ degrades the operational fidelity of the gated qubit pair $(Q_1,Q_3)$ in both OFF and ON configurations.  Full simulations that incorporate both decoherence and parasitism reveal an exponential growth in gate error for certain cells--most notably X and E1--as $\Gside$ increases, whereas other cells remain comparatively tolerant.

\begin{figure*}[t]
    \centering
    \includegraphics[width=0.99\linewidth]{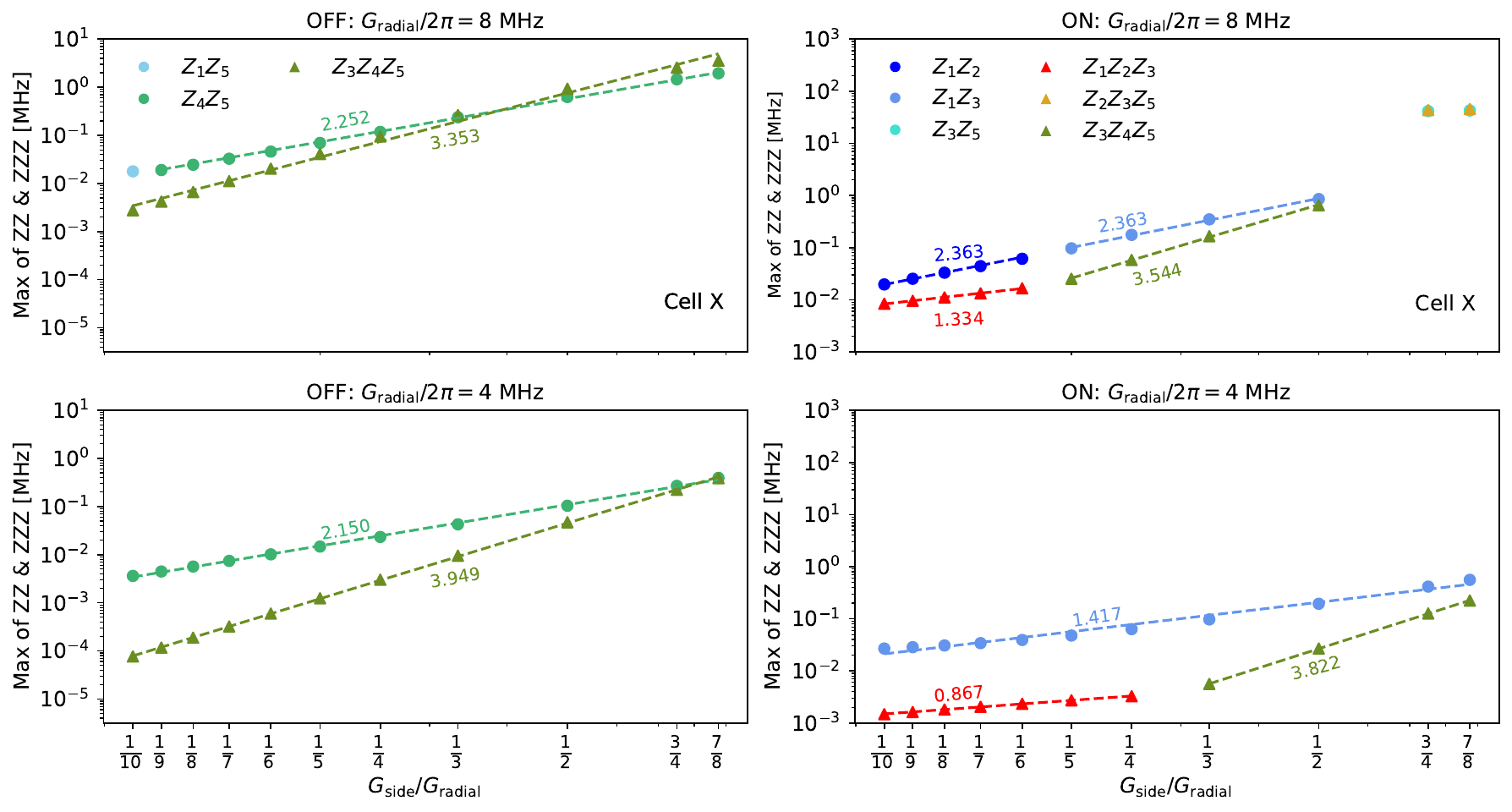}

\caption{Maximum parasitic interactions in Cell X against the coupling ratio $\Gside/\Grad$ for the OFF configuration (left) and ON configuration (right), evaluated at $\Grad/2\pi = 8~\text{MHz}$ (top panels) and $4~\text{MHz}$ (bottom panels).  Both axes employ logarithmic scales.  Each symbol marks the dominant parasitic channel at that point and represents a specific Pauli‐string interaction whose magnitude follows a power–law dependence on $\Gside/\Grad$; the fitted exponent is indicated alongside its trend line.  In the OFF state a ``phase transition''--signifying inversion of the interaction hierarchy, where the leading three‐body term overtakes the two‐body term--occurs at $\Gside/\Grad=1/2$ for the stronger coupling,  at $7/8$ for the weaker coupling.  In the ON state the same transition arises at $\Gside/\Grad=1/2$ for the strong‐coupling case, shifts beyond unity for the weak‐coupling case.}
    \label{fig:gside}
\end{figure*}

To probe this behavior more systematically, we examine the ratio of side   to radial  coupling, $\Gside/\Grad$.  We evaluated its impact on the gate fidelity of qubits $Q_1$ and $Q_3$ in both the ON and OFF states for two representative radial  strengths, $\Grad/2\pi = 4$ and $8~\mathrm{MHz}$.  For each case, we sweep $\Gside/\Grad$ up to unity and extract the maximal stray interactions $ZZ$ and $ZZZ$.  The resulting data for cell X appear in Fig.~\ref{fig:gside}, where the plot symbols identify the dominant parasitic component at each coupling ratio.  The upper panels correspond to the stronger case ($8~\mathrm{MHz}$) $\Grad$, while the lower panels refer to the weaker ($4~\mathrm{MHz}$) case.

\subsection*{Phase Transitions}

Quantum computing demands a Hamiltonian regime that preserves a clear hierarchy of many-body Pauli string interactions, with interaction strengths diminishing as the string length increases. Such a structure prevents the unwanted propagation of quantum states owing to residual many-body interactions during two-qubit gate operations.

Our numerical simulations, carried out with sufficiently large qubit arrays, confirm the existence of a robust quantum computational phase. However, this phase is inherently fragile, requiring precise control over circuit parameters and frequency assignments. Deviations from these optimal conditions may induce transitions into qualitatively different phases. Notably, one such regime exhibits an \emph{inverted} interaction hierarchy, characterized by weak direct two-qubit couplings overshadowed by dominant higher-order multiqubit interactions. This corresponds to a topologically ordered phase, extensively studied in condensed matter physics~\cite{KITAEV20032,WEN90,ansari2008statistical} and quantum information theory~\cite{RevModPhys.88.035005,zeng2015quantum,Nayak}. Encouragingly, our findings indicate that current hardware can experimentally access these topological phases through precise parameter tuning, although comprehensive studies remain necessary.

To expose the transition mechanism we vary the ratio of side‑qubit coupling to the primary (center–side) coupling, $\Gside/\Grad$, and track the largest matrix element in each Pauli sector. Figure~\ref{fig:gside} displays the results on a log–log scale: circles denote the dominant two‑body terms, triangles the dominant three‑body terms.
Interestingly, both follow a power-law trends with respect to the ratio $\Gside/\Grad$:
\begin{equation}
|ZZ|_{\mathrm{max}} \propto
\left(\frac{\Gside}{\Grad}\right)^{\ell_2},
\qquad
|ZZZ|_{\mathrm{max}} \propto
\left(\frac{\Gside}{\Grad}\right)^{\ell_3},
\label{eq:powerlaw}
\end{equation}
with exponents $\ell_2$ and $\ell_3$ that depend weakly on the unit‑cell geometry, gate state (ON/OFF), and absolute magnitude of $g_{\mathrm{rad}}$. Consistently across all cells we observe
\begin{equation}
\ell_2 \le \ell_3,
\label{eq:ells}
\end{equation}
indicating that three‑body interactions strengthen more rapidly than two‑body ones as $\Gside$ increases. Their curves intersect at a critical ratio $({\Gside}/{\Grad})^*$, signaling a crossover from the computational phase ($|ZZ|>|ZZZ|$) to the inverted, topologically ordered phase ($|ZZZ|>|ZZ|$).

The critical ratio $\left(\Gside/\Grad\right)^*$ is highly cell‑dependent and shifts between gate‑ON and gate‑OFF configurations. For example, when $\Grad/2\pi = 8\,\text{MHz}$ (top panels of Fig.~\ref{fig:gside}) the threshold moves from $1/3$ in the OFF state to $1/2$ in the ON state, effectively enlarging the computational phase. Other cells behave oppositely (see more details in Supplementary Fig. S4), underscoring the need to co‑optimize coupling strengths and gate schedules to guarantee robust fault‑tolerance throughout the lattice.

\subsection*{Logical Error}
Even if higher-order Pauli terms like $ZZZ$ are negligible at the physical level, they can accumulate and influence logical error rates in a quantum error-correcting code. 
For a distance-$d$ stabilizer code, the logical error probability scales approximately as
\begin{equation}
    p_L \sim C \, (p_{\mathrm{eff}})^{\lceil d/2 \rceil},
\end{equation}
where $p_{\mathrm{eff}}$ is the effective physical error rate including contributions from all relevant Pauli terms, and $C$ is a combinatorial factor accounting for the number of error patterns leading to a logical failure. 
If the physical error rate is $p_{\mathrm{phys}}$ and higher-order $k$-body terms have probabilities $p_k \sim \epsilon^k$, then
\begin{equation}
    p_{\mathrm{eff}} = p_{\mathrm{phys}} + \sum_{k>2} p_k.
\end{equation}
For instance, for a three-body term with $p_3 \sim 10^{-4}$ as shown in Fig. \ref{fig:iswap_fidelity4}(b) and a code distance $d=5$, the contribution to the logical error is
\begin{equation}
    p_L \sim (p_{\mathrm{phys}} + p_3)^{\lceil 5/2 \rceil} = (p_{\mathrm{phys}} + 10^{-4})^3.
\end{equation}
Assuming $p_{\mathrm{phys}} = 0.01$, we have
\begin{equation}
    p_L \approx (0.0101)^3 \approx 1.0303 \times 10^{-6},
\end{equation}
compared to $(0.01)^3 = 1 \times 10^{-6}$ without the higher-order term. 
This illustrates that even relatively small higher-order Pauli interactions can slightly increase logical error rates, and their cumulative effect becomes more significant for larger codes or near the fault-tolerance threshold, as the accumulation of many-body errors grows.

\section*{Discussion}
\label{sec:conclusion}

We have extended the QPU Hamiltonian formalism by introducing a systematic, diagrammatic approach for evaluating effective many-body Pauli interactions. Applied specifically to the five-qubit unit cell common in planar surface-code arrays, our method accurately captures both conventional two-body couplings and previously overlooked three-body terms arising from higher-order virtual processes. By integrating this compact diagrammatic representation with the comprehensive numerical diagonalization of full-circuit models, we mapped the evolution of parasitic couplings as functions of qubit detuning, coupler biases, and crucially the often neglected side-qubit coupling.

Our analysis reveals three distinct operational phases: (1) a low-error computational phase dominated by two-body interactions, ideal for scaling quantum processors; (2) an error-influenced yet still hierarchical phase, where three-body interactions are notable but subordinate; and (3) a hierarchy-inverted phase, in which three-body interactions surpass two-body terms, marking a transition into a topologically ordered regime with fundamentally altered error characteristics. Although the first two phases support quantum computation (albeit with increased errors in the second regime), the inverted hierarchy indicates a critical breakdown of conventional quantum processor operation. Our findings underscore that transitions between these phases can occur at the unit-cell scale, triggered by minor adjustments in gate parameters or qubit couplings.

From our detailed analysis, three key lessons emerge. First, if the side coupling exceeds approximately a limit, the standard interaction hierarchy can invert, elevating higher-order terms into the primary error source, a problem undetectable through standard two-qubit calibration procedures. Second, cell-specific behavior critically influences overall performance; thus, diagnostic tools like Processor Error Tomography (PET) are invaluable for promptly identifying and addressing high-risk unit cells before scaling up quantum circuits. Third, effective error mitigation must address both two-body and three-body interactions. Suppressing two-body terms alone is insufficient, as higher-order errors mediated through virtual couplings can remain significant. Targeted hardware solutions, such as optimizing wiring geometries to minimize side-qubit couplings, engineering flux-bias waveforms to suppress problematic frequency components, or employing tailored echo sequences, are essential to restore the intended interaction hierarchy.

Practically, this research advocates a parasite-aware approach to quantum processor design and calibration. Instead of attempting to eliminate side-qubit coupling, an often impractical goal, we demonstrate that maintaining a small direct coupling combined with optimized frequency assignments and coupler biases significantly enhances operational fidelity and reduces decoding overhead~\cite{xu24lattice,valles-sanclemente25optimizing}. This strategy promotes more consistent performance across processor tiles and simplifies the implementation of surface code protocols.

Looking ahead, our scalable QPU Hamiltonian framework readily accommodates systems comprising hundreds of unit cells and naturally extends to incorporate higher-order interactions, leakage phenomena, and tunable coupling noise. When coupled with experimental benchmarks from real quantum devices, this methodology provides a robust foundation for guiding the design and control strategies of next-generation processors, propelling the advancement toward fault-tolerant quantum computation on surface-code architectures.

\section*{Methods}

\subsection*{Hamiltonian Model}
In what follows, we recast the dynamics of a generic quantum processor into an \emph{effective Hamiltonian} acting on the \emph{near–computational} subspace, namely the logical manifold together with its adjacent leakage levels.  On this foundation, we develop a novel universal diagrammatic calculus that permits perturbative evaluation of every coefficient.  The prescription is scalable to the qubit number, lattice geometry, and Pauli string size, e.g. $ZZIII$, $IZIIZZI$, or $ZZIZIZZZZI$, and remains valid to all perturbative orders.  For clarity, we first illustrate the method on a five-qubit surface code unit cell before extending it to lattices of arbitrary size.

We adopt a qubit labeling convention in which each five-qubit surface code cell comprises a central transmon, $Q_{1}$, surrounded by four transmons $Q_{2}$, $Q_{3}$, $Q_{4}$ and $Q_{5}$ (see Fig.~\ref{fig:unit_cell}). The \emph{Sycamore} processor architecture employs tunable couplers that suppress crosstalk during idle states while selectively activating interactions during computation.

Each unit cell hosts four frequency-tunable couplers $C_{1i}$, which mediate interactions between the central qubit $Q_{1}$ and its adjacent qubits $Q_{i}$ at the corners, with $i=2,3,4,5$.
The corresponding coupling strength, denoted $G_{Q_{i}C_{1i}}$, is obtained by mapping the circuit Lagrangian to its Hamiltonian, thereby exposing the dependence on circuit parameters that govern phase and charge exchange between couplers and qubits.
Furthermore, we account for residual ``side couplings'' $G_{Q_{i}Q_{j}}$ between a pair of qubits, as well as stray capacitive interactions between the central and side qubits, referred to as ``radial couplings'' $\Grad$. Coupler frequency tuning enables a broad adjustment of qubit-qubit interactions from negative to positive values. All couplings are mandated to be symmetric under switching indices by reciprocity in superconducting circuits.

A rough estimate of the two-body interaction Hamiltonian in qubit-only lattices has long been obtained---either perturbatively \cite{devoret2013superconducting,koch2007charge-insensitive,houck2012on-chip,blais2004cavity} or non-perturbatively \cite{cederbaum1989block,ansari19superconducting,ansari0method}. Projecting onto the subspace in which the couplers remain in their ground states then yields renormalized qubit frequencies ($\tilde{\omega}_{Q}$), anharmonicities ($\tilde{\delta}_{Q}$), and effective qubit–qubit interactions.

\begin{figure*}[t]
    \centering
    \includegraphics[width=0.8\linewidth]{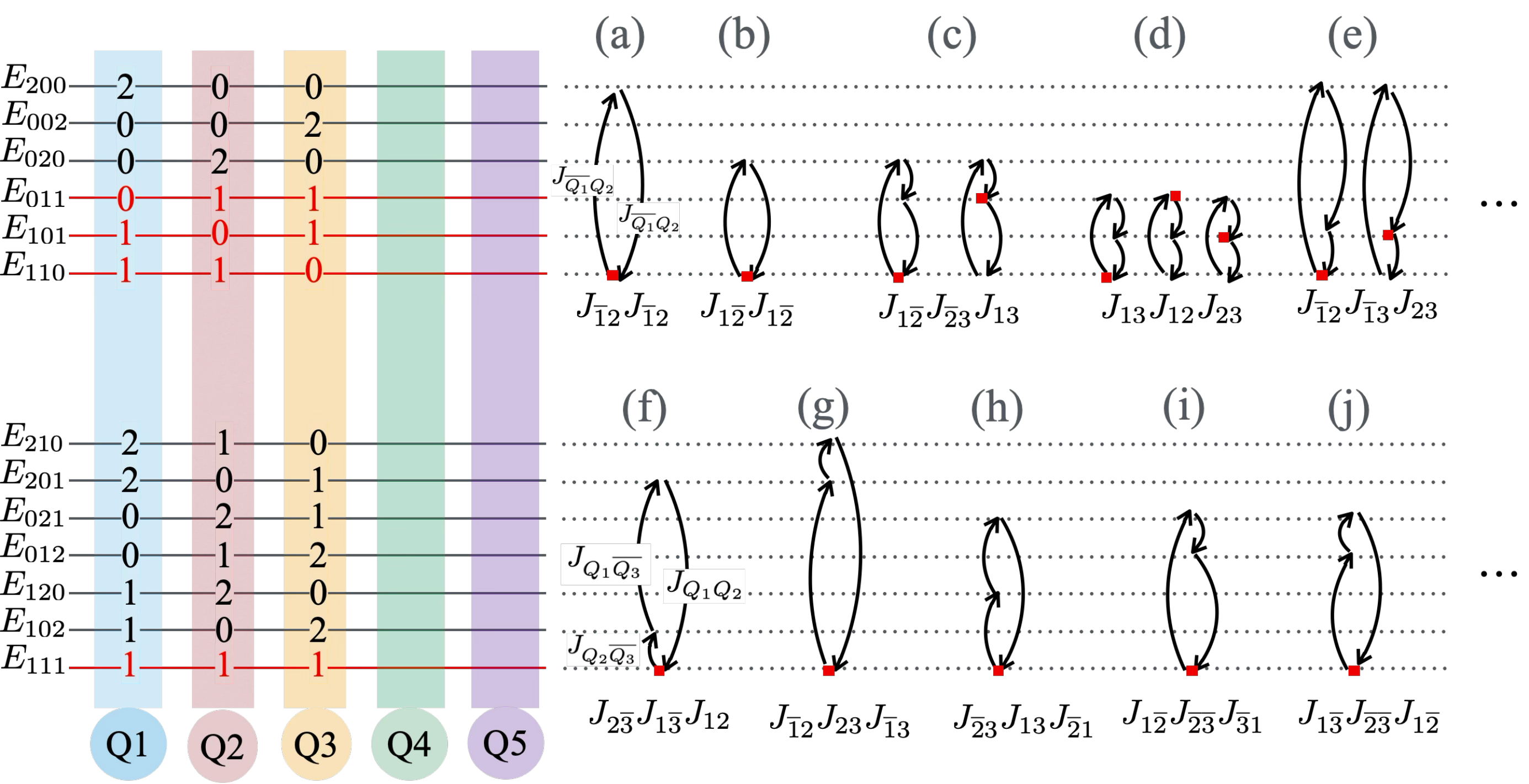}
\caption{Typical diagrams for the perturbative evaluation of the Pauli interaction term $ZZI$ in the three-qubit basis. Left: Computational states are highlighted in red within the two-excitation (top) and three-excitation (bottom) manifolds. Right: Panels (a–e) illustrate typical transitions between marked two-excitation computational and non-computational levels, while panels (f–j) show the analogous three-excitation processes. Diagrams that differ only in their initial marked state are grouped together, with the red dot indicating the marked level that distinguishes each group. }\label{fig:jdiag}
\end{figure*}
 
The effective qubit-qubit interaction, $J_{ij}$, emerges from virtual tunneling processes where an excitation briefly migrates from qubit $Q_i$ to a coupler and subsequently to qubit $Q_j$ or vice versa. Exchanges can occur among computational levels ($|0\rangle$, $|1\rangle$), between computational and higher non-computational levels ($|2\rangle$, $|3\rangle$, etc.), or exclusively among higher-level states. We explicitly indicate these states through indices: interactions between computational states as $J_{Q_iQ_j}$, higher-level states as $J_{\overline{Q_i}\overline{Q_j}}$, and mixed states as $J_{Q_i\overline{Q_j}}$. Sometimes we simplify the notation in this text by removing ``Q'' and only mentioning the qubit label. Additionally, transitions involving the $|2\rangle \leftrightarrow |1\rangle$ states require a prefactor $\sqrt{2}$ in front of the corresponding $J$, reflecting ladder operator definitions. A generalization for the third excited level includes double overlines and a prefactor $\sqrt{3}$, e.g., $\sqrt{3} J_{\overline{\overline{Q_i}}Q_j}$.

The perturbative analysis yields the following approximate relationship for the coupling between any pair of qubits, say $Q_p$ and $Q_q$, within computational levels: 
\begin{equation}
    J_{pq} \approx G_{Q_pQ_q} + \frac{{{G_{Q_p C_{pq}} G_{Q_q C_{pq}}}}}{2}\left(\frac{1}{\Delta_p}+\frac{1}{\Delta_q}\right), 
    \label{eq. pert J}
\end{equation}
with frequency detuning between the qubit $Q_p$ and the coupler $C_{pq}$ being $\Delta_p=\omega_{C_{pq}} - \omega_{Q_p}$.

Beyond the dispersive regime ($\Delta \gg |G|$), perturbative evaluations lose accuracy. 
To address this, we have developed a non-perturbative method, referred to as CirQubit~\cite{cirqubit}, which leverages minimal alterations to the Hamiltonian spectrum \cite{cederbaum1989block,magesan20effective,xu2021zz-freedom}. 
In this approach, off-diagonal terms corresponding to qubit–coupler interactions are systematically removed in the combined qubit–coupler basis, truncated by an excitation overhead, while the best matching between eigenvalues and eigenvectors is tracked and the original ordering of the Hamiltonian is preserved during diagonalization. Due to the one-to-one eigenvalue and eigenvector matching, this approach remains reliable in the inverted regime, where the coupler frequency approaches the qubit frequencies, leading to strong qubit-qubit coupling and then many-body localization theory fails~\cite{berke2022transmon}. The effective resulting Hamiltonian $H_{\text{eff}}$ directly reveals the interaction parameters $J$, which guide the precise engineering of qubit interactions.

By eliminating virtual coupler states, the idle circuit Hamiltonian is confined exclusively to qubit-qubit interactions. The computational processor Hamiltonian after diagonalization thus reads:
\begin{equation}
H = \sum_{i} \alpha_i \hat{Z}_i + \sum_{i,j} \alpha_{ij} \hat{Z}_i \hat{Z}_j 
+ \sum_{i,j,k} \alpha_{ijk} \hat{Z}_i \hat{Z}_{j} \hat{Z}_k + \cdots 
\label{eq. H}
\end{equation}
with $\alpha$'s representing the strength of each Pauli-string interaction. 

\subsubsection*{Diagrammatic Approach to perturbative $H$}

Evaluation of parasitic interactions using perturbation theory or the Schrieffer-Wolff transformation typically requires a perturbative analysis of each computational state, and higher-order corrections, e.g., up to third order, involve significant computational effort. These approaches can become cumbersome and computationally expensive, especially as the number of qubits increases, and provide limited intuition about the origin and control of parasitic interactions.
To address these limitations, we propose a diagrammatic framework that evaluates parasitic interactions more efficiently and directly in the dispersive regime. By using a simple summation scheme for each diagram, with each perturbation order represented by the number of trajectories in the loop, only interactions that contribute to energy shifts are considered, avoiding unnecessary complexity. The diagrams also account for discrepancies in the $J$ couplings between different levels, marked by new notations, which is nontrivial but essential for accurate estimation. Importantly, the framework is fully scalable to higher orders and to an arbitrary number of qubits, making it both efficient and broadly adaptable.

To illustrate, consider the three-body Pauli term $ZZI$ in a five-qubit configuration, as shown in Fig.~\ref{fig:jdiag}. The corresponding Pauli operator applies the $Z$ operator exclusively to qubits $Q_1$ and $Q_2$.

A representative two-qubit Pauli string---for instance $ZZIII$---acquires an effective coupling
\begin{equation}
\sum_{n_3,n_4,n_5\in{\{0,1\}}}^{(\geq2)}
\hspace{-0.6cm}E_{11 n_3 n_4 n_5}+E_{00 n_3 n_4 n_5}-E_{10 n_3 n_4 n_5}-E_{01 n_3 n_4 n_5},
\end{equation}
where the superscript $(\geq2)$ confines the sum to manifolds containing at least two excitations---e.g.\ $E_{11000}$ or $E_{10001}$---so that all single-excitation contributions cancel. This definition can be rewritten as a summation over all individual energy levels subject to a \emph{parity rule} that multiplies some terms by a negative cofactor.  

$\blacktriangleright$ \emph{Parity rule--}Attach a factor $(-1)$ to each qubit that carries a $Z$ and occupies $|1\rangle$. For the string $ZZIII$ this yields
\begin{align*}
 &   (-1)^2E_{11 n_3 n_4 n_5}+(-1)^0E_{00 n_3 n_4 n_5}+ (-1)^1E_{10 n_3 n_4 n_5}+(-1)^1E_{01 n_3 n_4 n_5},
\end{align*}

In general, evaluating parasitic interactions requires considering all computational states with more than one excitation, which we refer to as \emph{marked levels}. For example, in an $n$-qubit basis, evaluating $ZZ$ or $ZZZ$ interactions requires considering $2^n - n - 1$ computational states. For the $ZZI$ and $ZZZ$ terms in the three-qubit basis, the four computational eigenstates $|110\rangle$, $|101\rangle$, $|011\rangle$, and $|111\rangle$ that define the Pauli term constitute the marked levels. Figure~\ref{fig:jdiag} highlights them in red on the right; virtual couplings between these levels and the surrounding spectrum set the Pauli-string strength with arrows representing the transverse couplings.

Consider an $N$-qubit Pauli term in the general form
$\prod_{i=1}^{N}\hat{O}_{i}$, where each operator $\hat{O}_{i}$ acting on qubit $Q_{i}$ is either the Pauli–$Z$ operator $\hat{Z}$ or the identity $\hat{I}$. This term is evaluated as a weighted sum over computational-basis states that may contain multiple excitations. To determine the sign of the contribution from a specific energy level $E_{n_{1}n_{2}\cdots n_{N}}$, where the binary variable $n_{i}\in{0,1}$ denotes the excitation state of qubit $Q_{i}$, we employ the following rule. For each qubit $Q_{i}$ that appears with a $Z$ operator ($\hat{O}{i}=\hat{Z}$), assign a factor of $-1$ if $n_{i}=1$. Consequently, the contribution of that energy level to the Pauli string acquires the overall sign
$$
\prod_{i=1}^{N}(-1)^{\delta_{\hat{O}_{i},\hat{Z}}\times \delta{n_{i},1}},
$$
where the Kronecker delta $\delta_{A,B}$ equals $1$ when $A=B$ and $0$ otherwise, for either c-number or operator arguments $A$ and $B$. The overall sign of the contribution from any marked level is therefore the product of all such $-1$ factors. This parity rule offers a simple and universally applicable method for assigning correct signs when summing over marked computational states in the perturbative evaluation of any given Pauli term.$\blacktriangleleft$

To determine the strength of a Pauli string in the lattice Hamiltonian, we first identify the {marked} computational levels together with any nearby non-computational (unmarked) levels that share the same total excitation number. Within this diagrammatic formalism, evaluating a Pauli term then amounts to analyzing a set of individual diagrams. Each diagram traces single-particle virtual transitions that originate from a marked level, pass through $p$ intermediate (computational or non-computational) states, and ultimately return to the same marked level. Such a pathway contributes at $p^{\text{th}}$ order to the Pauli strength. Because every hop carries the coupling constant $J$, the amplitude of the diagram scales as $J^{p}$. Formally, the contribution equals the product of all qubit–qubit transition amplitudes from the marked state through the intermediate states and back again. This product is divided by the product of the energy gaps $\Delta$ that separate the marked state from each intermediate state encountered, yielding the factor $J^{p}/\Delta^{p-1}$. 

The procedure applies uniformly to every diagram in the expansion. As an illustrative example, we consider the Pauli term $ZZI$, whose diagrams up to third order are shown in Fig.~\ref{fig:jdiag}. Diagrams (a) and (b) correspond to second-order processes, while diagrams (c)–(j) represent third-order contributions. In particular, diagram~(d) can be decomposed into three analogous subdiagrams that differ only in the choice of computational level involved, and their amplitudes are summed according to the parity rule. As a concrete illustration, explicit algebraic expressions for diagrams~(d), (e), (g), and~(i) in Fig.~\ref{fig:jdiag} are presented below.

\begin{align}
\text{Diagram (d): } & 2 J_{12}J_{23}J_{13}\left[\frac{1}{(E_{110}-E_{011})(E_{110}-E_{101})} -\frac{1}{(E_{011}-E_{101})(E_{011}-E_{110})} \right.\nonumber\\
&\left.-\frac{1}{(E_{101}-E_{110})(E_{101}-E_{011})}\right],\nonumber\\ 
\text{Diagram (e): } & 4 J_{\overline{1}2}J_{23}J_{\overline{1}3}\left[\frac{1}{(E_{110}-E_{200})(E_{110}-E_{011})} -\frac{1}{(E_{101}-E_{110})(E_{101}-E_{200})}\right],\nonumber\\ 
\text{Diagram (g): } & \frac{4J_{\overline{1}2}J_{23}J_{\overline{1}3}}{(E_{111}-E_{201})(E_{111}-E_{210})},\nonumber \quad\text{Diagram (i): } \frac{8J_{1\overline{2}}J_{\overline{23}}J_{1\overline{3}}}{(E_{111}-E_{021})(E_{111}-E_{012})}.\nonumber
\end{align}
\vspace{-0.2cm}

Note that in these diagrams, one can go from any marked state to intermediate state in the opposite direction sketched by arrows; therefore, we multiplied each diagram by a factor of 2.

Before summarizing the diagrammatic rules for evaluating a quantum-processing unit’s circuit Hamiltonian, we introduce a compact notation that streamlines the energy denominators appearing in each diagram. In any such diagram, the denominator is the product of the energy differences between the marked level and every intermediate level. For example,
$$
E_{110}-E_{020}
  =\bigl(E^{Q_1}_{1}-E^{Q_1}_{0}\bigr)
  +\bigl(E^{Q_2}_{1}-E^{Q_2}_{2}\bigr)
  =f^{Q_1}_{1\to0}-f^{Q_2}_{2\to1},
$$
where $f^{Q_i}_{m\to n}$ ($m>n$) denotes the transition frequency from level $m$ to level $n$ of qubit $Q_i$ (assume $\hbar=1$).

\medskip
\paragraph*{Squeezed–denominator symbol:} We define
$$
\Delta_{\overline{Q_i}Q_j}=f^{Q_i}_{2\to1}-f^{Q_j}_{1\to0},
\qquad\text{(abbreviated }\Delta_{\overline{i}j}\text{).}
$$
An index without an overline refers to the $|1\rangle\leftrightarrow|0\rangle$ transition; a single overline refers to the $|2\rangle\leftrightarrow|1\rangle$ transition; a double overline to the $|3\rangle\leftrightarrow|2\rangle$ transition, and so on. Consequently,
$$
E_{110}-E_{020}=\Delta_{Q_1\overline{Q_2}}.
$$

\noindent
This notation “squeezes’’ multi-term energy differences into a single, easily readable symbol, thereby greatly simplifying higher-order diagrammatic expressions.

\subsubsection*{Summary of Diagrammatic Rules}

We now summarize what we discussed above into a set of rules that help to evaluate perturbative estimation of all terms in the lattice Hamiltonian of a processor.  These rules are valid within the dispersive regime, where $J / \Delta < 1$, and can be used for arbitrary Pauli strings of any number of qubits (e.g. $ZZIII$, $IZIIZZI$, $ZZIZIZZZZI$) for any perturbation orders.

For a given processor of $N$ qubits, labeled $Q_i$ with $i=1,\cdots, N$, the lattice Hamiltonian after removing all the couplers and simplifications in the diagonal frame can be written in the format of Eq.~\eqref{eq. H} by replacing the upper qubit label from 5 to $N$. The $\alpha$'s are the coefficients of the Pauli operator terms $\bigotimes_{i=1}^N \hat{O}_i$ with $\hat{O_i}={\hat{I},\hat{Z}}$ being the operator that acts on the qubit $Q_i$.  For evaluating a typical term like the Pauli coefficient in the perturbative order $p$ we use the following diagram rules:

\begin{enumerate}
\item Mark the initial computational energy level (red dot).
\item Identify intermediate computational and non-computational levels reachable by exchanging $p$ particles.
\item Repeat for each marked initial level.
\end{enumerate}

The steps above prepare the diagrams for evaluation of their many-body Pauli-string  coefficients. Consider for any given diagram, one can take the following steps to evaluate the diagram:

\begin{enumerate}
\item Assign $J$ coefficients with appropriate overlines and prefactors based on the excitation levels involved.
\item Determine the sign of each diagram using the generalized parity rule.
\item Divide each diagram by the product of energy gaps between the marked initial level and all intermediate levels.
\item Note that different orderings of intermediate transitions may yield different diagrams. Each such diagram should be treated independently and included in the final sum, with signs assigned according to the parity rule.
\end{enumerate}

The diagrammatic framework developed here is not restricted to superconducting qubit architectures and can, in principle, be extended to spin-qubit systems. In such platforms, effective diagonal interactions arise from the perturbative elimination of excited orbital, valley, or spin states mediated by exchange and spin–orbit couplings. These off-diagonal processes generate energy shifts within the computational subspace, leading to effective $ZZ$ and higher-order multi-qubit interactions that can be organized diagrammatically in direct analogy to the superconducting-qubit case. While the specific microscopic mechanisms and dominant coupling channels differ, the essential perturbative structure remains similar, suggesting that the framework provides a general approach for analyzing parasitic interactions across diverse qubit technologies.

\subsection*{Numerical simulation of the iSWAP gate}
For every unit cell, we estimate the gate fidelity via quantum process tomography, employing device-reported coherence times at idle ($T_2 = T_1$).  The decoherence effects are included by solving the Lindblad master equation numerically. 
At each time step, the density matrix is modeled as
\begin{equation}
\rho_{t+1} =
\Lambda_{T}(Q_1)\!\circ\!\Lambda_{T}(Q_3)\!\circ\!
U_{\text{stray}}\!\circ\!
U_{\text{iSWAP}}[\rho_{t}],
\end{equation}
with the stray unitary evolution being 
\begin{equation}
\label{eq.Ustray}
U_{\text{stray}}
  =U\!\bigl(Z_1Z_3,\,Z_1Z_2Z_3,\,Z_1Z_3Z_4,\,Z_1Z_3Z_5\bigr)
 \end{equation}
 and the iSWAP gate unitary operation being 
  \begin{equation}
U_{\text{iSWAP}}
  =\exp \bigl[-iHt\bigr],\;
H=\tfrac{J_{13}}{2}(X_1X_3+Y_1Y_3).
\label{eq.Uiswap}
\end{equation}

\section*{Acknowledgement} 
Not Applicable
\section*{Author Contributions}
M.A., J.M., and X.X. conceived the idea. Detailed analysis was carried out by X.X. with input from M.A., K.K. and C.V. contributed to the theoretical derivations, and C.V. prepared several figures. X.X. wrote the manuscript with input from M.A., C.V., and J.M., M.A. and J.M. supervised the project. All authors reviewed the manuscript.

\section*{Competing interests}
The authors declare no competing interests.

\section*{Funding} 
 This research received funding from the Horizon Europe OpenSuperQPlus100 project (Grant Agreement No. 101113946).

\section{Unit cell Hamiltonian}\label{App.H_cell}
\label{app:sec.pertZZandZZZ}

Given a diamond-like surface code where one central transmon $Q_1$ couples to four side qubits $Q_2-Q_5$ counterclockwise via both indirect tunable couplers and direct capacitances, the corresponding Hamiltonian is expressed as 
\begin{align}
H =& \sum_{i=1}^{5} \omega_{Q_i} b_i^\dagger b_i 
+ \frac{\delta_i}{2} b_i^\dagger b_i (b_i^\dagger b_i - 1)+ \sum_{i=2}^{5} \omega_{C_{1i}} a_i^\dagger a_i  +\sum_{i=2}^{5} \Grad (b_1 - b_1^\dagger)(b_i - b_i^\dagger)\nonumber\\
&  + \sum_{i=2}^{5} G_{Q_iC_{1i}} (b_i - b_i^\dagger)(a_i - a_i^\dagger) + \sum_{i=2}^{5} G_{\text{side}} (b_i - b_i^\dagger)(b_{i+1} - b_{i+1}^\dagger)
\end{align}

In the dispersive regime where $|G/(\omega_{C_{1i}}-\omega_i)|\ll 1$ such 5-qubit unit cell is reduced to the well-established effective Hamiltonian in the multilevel qubit basis:  
\begin{align}\label{eq:Hj}  
H_{q}=&\sum_{i, n_i}\bar{E}_{Q_i}({n_i})|n_i\rangle\langle n_i| +\sum_{i,j, m_{i},n_j} \sqrt{(m_{p}+1)(n_{q}+1)}  \nonumber\\  
& \times \mathcal{J}_{ij} (|m_{i}, n_j+1\rangle\langle m_{i}+1,n_j|+\text {H.c.}),  
\end{align}  
where the dressed qubit frequency is $\bar{\omega}_{Q_i}(n_i) = \bar{E}_{Q_i}(n_i+1) - \bar{E}_{Q_i}(n_i)$, with $\bar{E}_{Q_i}(n_i)$ being dressed state energies. $\mathcal{J}_{ij}$ is the abstract symbol for all level-dependent interactions between qubit $i$ and $j$, i.e. $\mathcal{J}_{ij}=\{J_{ij}, J_{\overline{i}j}, J_{i\overline{j}}, J_{\overline{\overline{i}}j}, J_{\overline{i}\overline{j}}, \cdots\}$ and as discussed in the main text, no overline refers to $0\leftrightarrow 1$ transition, one overline for $1\leftrightarrow 2$, two overlines for $2\leftrightarrow 3$, etc. Within the computational subset the qubit-qubit interaction can be approximated to $J_{ij} \approx G_{Q_iQ_j} + (1/2){G_{Q_i C_{1i}} G_{Q_j C_{1j}}}\left({1}/{\Delta_i}+{1}/{\Delta_j}\right)$, with $\Delta_i=\omega_{C_{1i}} - \omega_{Q_i}(n_i)$. Similarly, the detuning is also defined with an overline if the interaction occurs between computational and non-computational subspaces.

The \textsc{Cirqubit}~\cite{cirqubit} software automatically computes the system Hamiltonian in the hybrid basis of qubits and couplers using a nonperturbative approach. The full Hamiltonian is first block diagonalized via a modified principle of least action, combined with a non-repetitive eigenvector sorting procedure designed to decouple the coupler degrees of freedom. From the resulting block structure, the effective qubit-qubit exchange rate $J$ is extracted. The reduced Hamiltonian is then fully diagonalized to evaluate residual parasitic interactions, including multi-body coupling terms.

\section{Examples of bad cell}\label{app:bad_cell}

\begin{figure*}[ht]
\centering
\includegraphics[width=0.95\textwidth]{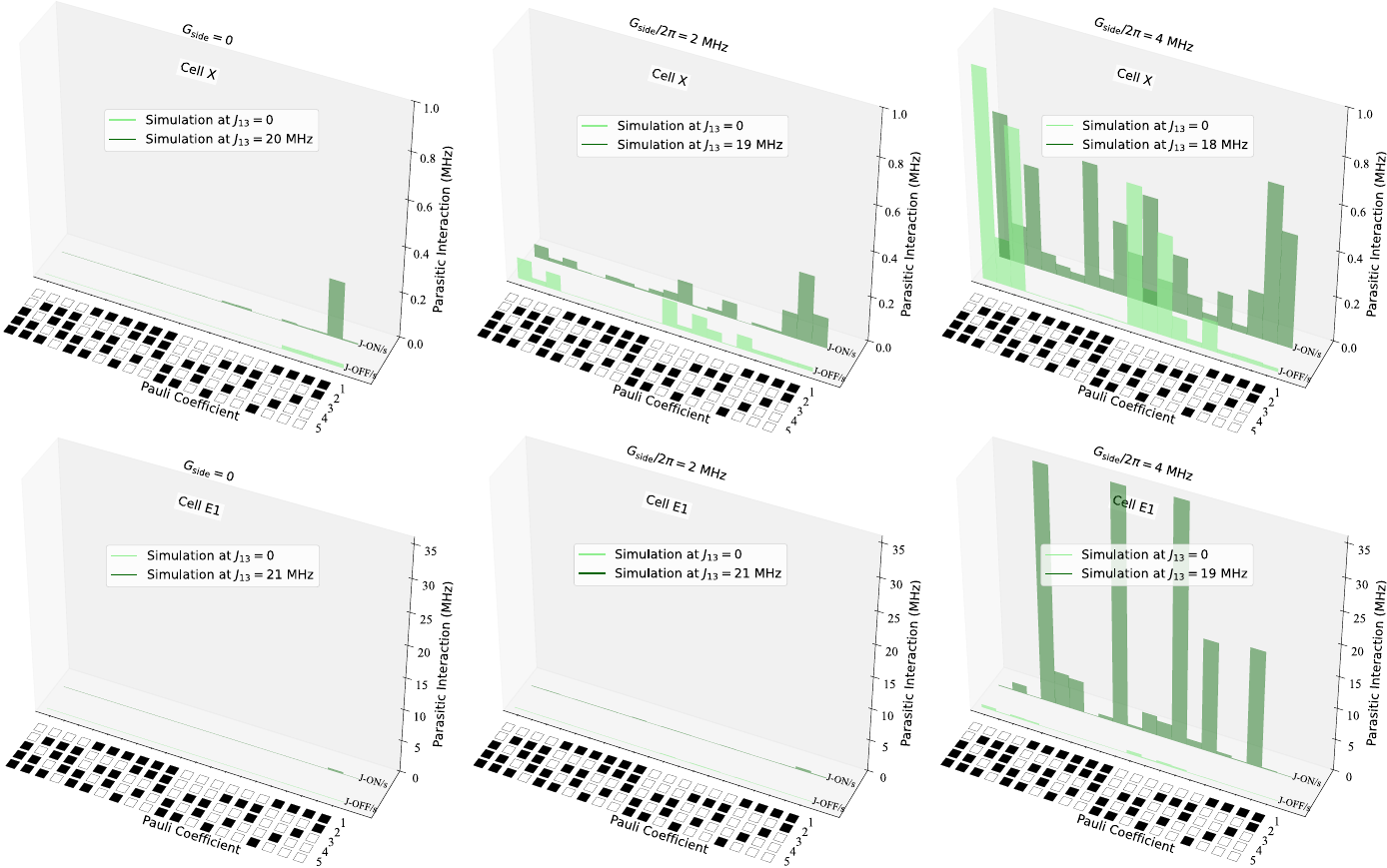} \put(-355,225){\textbf{(a)}}\put(-235,225){\textbf{(b)}}\put(-115,225){\textbf{(c)}}
 \put(-355,115){\textbf{(d)}}\put(-235,115){\textbf{(e)}}\put(-115,115){\textbf{(f)}}

\caption{Hamiltonian tomography of parasitic interactions for both the effectively \emph{off} and \emph{on} states of unit cells X (\textbf{top}) and E1 (\textbf{bottom}). Light-colored bars indicate $J_{\text{OFF}}$, while deep-colored bars indicate $J_{\text{ON}}$. The x-axis represents 20 different parasitic interactions, where two- and three-qubit Pauli-$Z$ terms are filled in black, and Pauli-$I$ terms are left unfilled. The parasitic interactions are evaluated under three scenarios: (a) and (d) $G_{\text{side}} = 0$, (b) and (e) $G_{\text{side}}/2\pi = 2~\text{MHz}$, and (c) and (f) $G_{\text{side}}/2\pi = 4~\text{MHz}$, with results obtained from numerical simulations using \textsc{Cirqubit}.}
\label{fig:bad_cells}
\end{figure*}

We evaluated the optimized iSWAP gate fidelity across all unit cells and found that cells X and E1 exhibit noticeable fidelity degradation. To investigate the underlying cause, we computed all parasitic interactions for these two cells at different values of $G_{\text{side}}/2\pi = 0$, $2$, and $4$~MHz, as shown in Fig.~\ref{fig:bad_cells}.

The relevant parasitic interactions are those involving qubits $Q_1$ and $Q_3$, namely $Z_1Z_3$, $Z_1Z_2Z_3$, $Z_1Z_3Z_4$, and $Z_1Z_3Z_5$. The top panels of Fig.~\ref{fig:bad_cells} display the parasitic couplings on unit cell X. It is evident that at $G_{\text{side}}/2\pi = 4$~MHz, the interactions $Z_1Z_3$ and $Z_1Z_3Z_4$ exceed 100~kHz, resulting in a noticeable degradation in iSWAP fidelity. In the bottom panels for unit cell E1 at the same side coupling strength, three-body interactions--particularly $Z_1Z_3Z_5$--dominate, leading to a significant fidelity drop between the decoherence $+$ $ZZ$ + $ZZZ$ and decoherence $+$ $ZZ$ models.

\begin{figure}[ht]
\centering
            \includegraphics[width=0.65\linewidth]{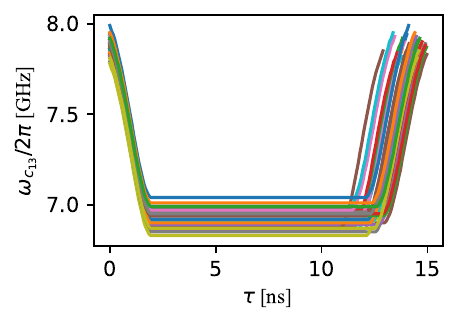}
\caption{Flat-top Gaussian pulse shapes of all 33 unit cells.}
\label{fig:pulse}
\end{figure}

The pulse shaping used in the simulation consists of a flat-top pulse with two rapid 2-ns Gaussian rise and fall edges on either side, as shown in Fig.~\ref{fig:pulse}. 

\begin{figure*}[ht]
    \centering
    \includegraphics[width=0.9\textwidth]{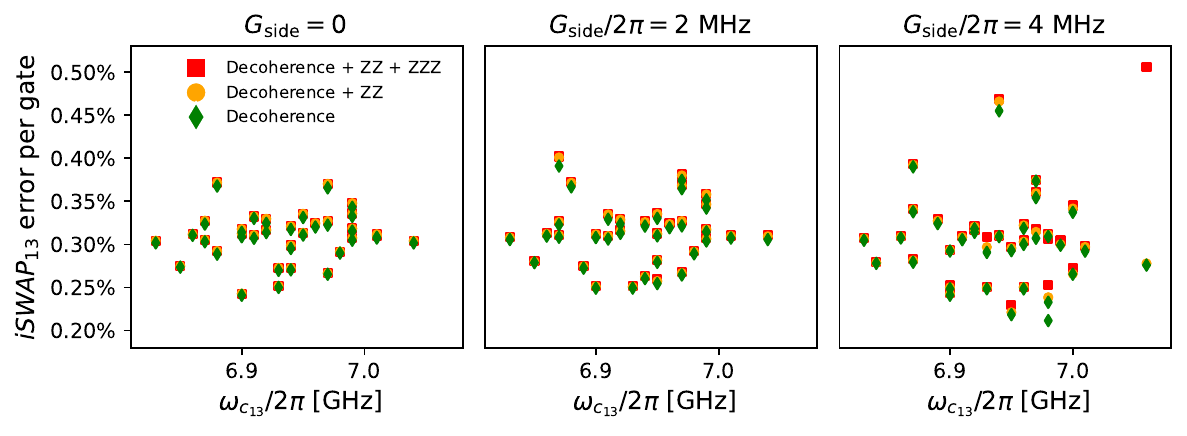}
   \caption{Classified iSWAP error rate of all unit cells versus coupler frequencies at $\Gside/2\pi = 0$, $2$, and $4$ MHz, shown separately.}
    \label{fig:popfid}
\end{figure*}

Before we conclude this section, let us briefly indicate the impact of these capacitive couplings such as $\Gside$ and $\Grad$ on the fidelity of gate acting on active qubits.  We consider the optimum fidelity performance obtained in the main text and replot them classified in three sectors of $\Gside/\Grad$ being 0, $1/4$ and $1/2$---corresponding to $\Gside=0,2,4$ MHz with $\Grad=8$ MHz. The results are shown in Fig.~\ref{fig:popfid} where the gate fidelity of all cells is plotted versus their corresponding coupler frequency. These plots show that by the increase of side-qubit couplings, errors increase not only for different scenarios indicated in the legend but also the data points are spread across a wider range of coupler frequencies on the x-axis.   

\section{Second example of phase transitions}\label{app:phase_cellw}

This section presents another example, focusing on Cell W, to illustrate the impact of the ratio between side   coupling and radial  coupling, $\Gside / \Grad$.

\begin{figure*}[ht]
    \centering
    \includegraphics[width=0.9\linewidth]{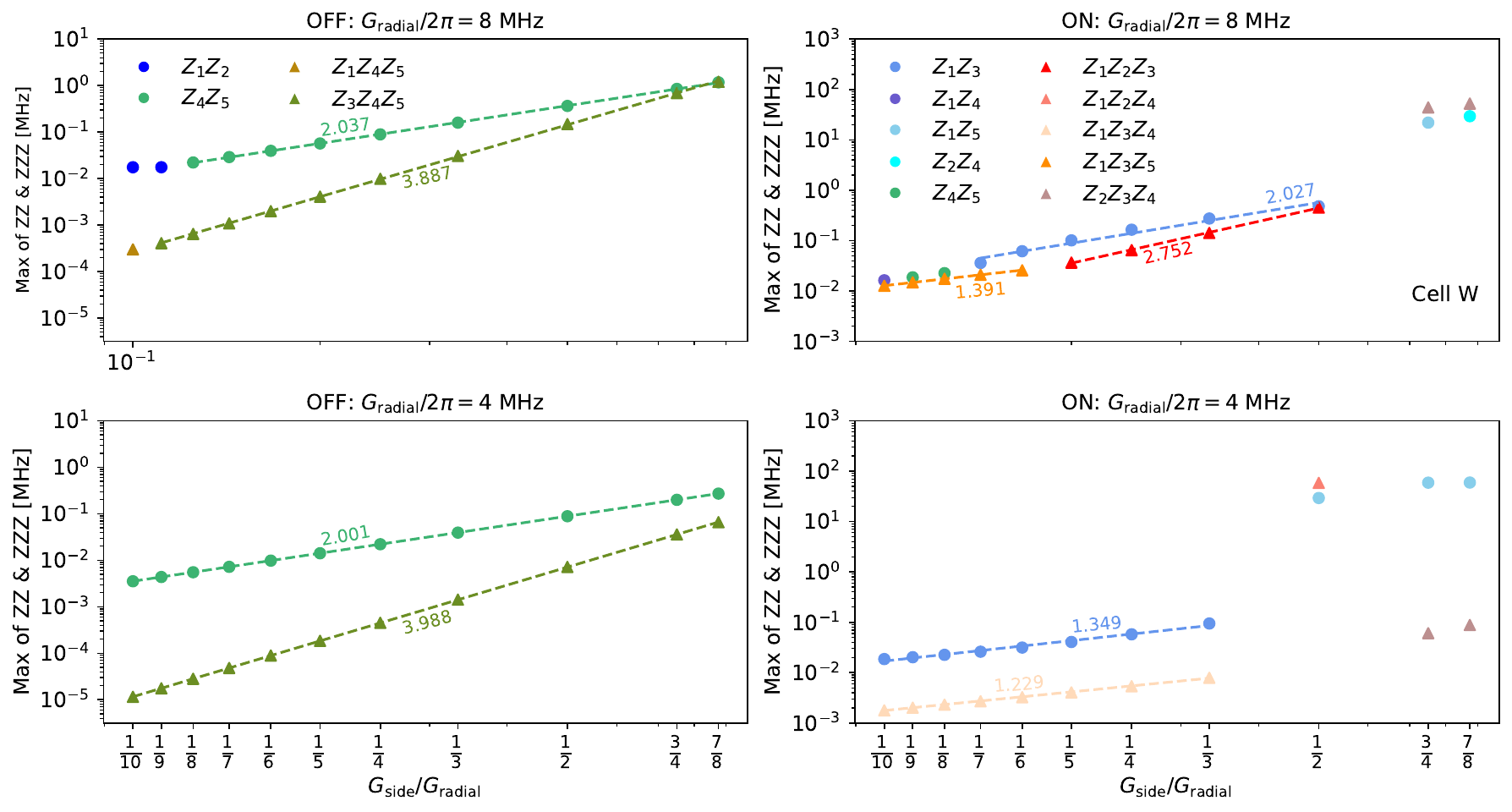}\put(-335,180){\textbf{(a)}}\put(-170,180){\textbf{(b)}}\put(-335,92){\textbf{(c)}}\put(-170,92){\textbf{(d)}}
    \caption{Maximum parasitic interactions in Cell W against the coupling ratio $\Gside/\Grad$ for the OFF configuration (left) and ON configuration (right), evaluated at $\Grad/2\pi = 8~\text{MHz}$ (top panels) and $4~\text{MHz}$ (bottom panels).  Both axes employ logarithmic scales.  Each symbol marks the dominant parasitic channel at that point and represents a specific Pauli‐string interaction whose magnitude follows a power–law dependence on $\Gside/\Grad$; the fitted exponent is indicated alongside its trend line.  In the OFF state a ``phase transition''--signifying inversion of the interaction hierarchy, where the leading three‐body term overtakes the two‐body term--occurs at $\Gside/\Grad=7/8$ for the stronger coupling,  shifts beyond unity for the weak‐coupling case.  In the ON state the same transition arises at $\Gside/\Grad=1/2$ for the strong‐coupling case, shifts beyond unity for the weak‐coupling case.} \label{fig:gsidew}
\end{figure*}

\paragraph*{OFF State Behavior:} 
Similar to Cell X, the maximum stray interactions increase monotonically with $\Gside$. As shown in Fig.~\ref{fig:gsidew}(a), the dominant errors originate from $Z_4Z_5$ and $Z_3Z_4Z_5$. The interaction strength exhibits slight dependence on the coupling strength $\Grad$; in the weak (strong) coupling regime, the extracted powers are $\ell_2 = 2.001\,(2.037)$ and $\ell_3 = 3.988\,(3.887)$ in the OFF state. Notably, when $\Gside/\Grad \approx 3/4$ at $\Grad/2\pi = 4~\mathrm{MHz}$, or $\approx 1/3$ at $\Grad/2\pi = 8~\mathrm{MHz}$, i.e., when $\Gside/2\pi \simeq 2~\mathrm{MHz}$, the two-body parasitic terms increase to several hundred kilohertz--still potentially tolerable for gate fidelity. A phase transition occurs at $\Gside/\Grad = 7/8$ in the strong-coupling case, shifting beyond unity under weaker coupling. The phenomenon of \emph{$ZZZ$-superiority}, where three-body terms dominate two-body interactions, emerges only at higher $\Gside$ and persists even at lower side couplings.

\paragraph*{ON State Behavior ($\Grad/2\pi = 4~\mathrm{MHz}$):} 
At low side couplings, the main two-body parasitics are $Z_1Z_3$ and $Z_1Z_3Z_4$. As $\Gside$ increases, the dominant three-body term transitions from $Z_1Z_3Z_4$ to $Z_1Z_2Z_4$, and subsequently to $Z_2Z_3Z_4$, as illustrated in Fig.~\ref{fig:gsidew}(b). A phase transition occurs at $\Gside/\Grad = 1/2$ in the strong-coupling regime and shifts beyond unity for weaker couplings.

\paragraph*{ON State Behavior ($\Grad/2\pi = 8~\mathrm{MHz}$):} 
With stronger direct coupling, the migration of dominant interactions is more pronounced. The leading two-body term shifts in the sequence $Z_1Z_4 \rightarrow Z_4Z_5 \rightarrow Z_1Z_3 \rightarrow Z_2Z_5 \rightarrow Z_2Z_4$, while the leading three-body term follows $Z_1Z_3Z_5 \rightarrow Z_1Z_2Z_3 \rightarrow Z_2Z_3Z_4$. The phase transition similarly occurs around $\Gside/\Grad \approx 1/2$ under strong side coupling. This cascading evolution underscores the increasing significance of side-qubit interactions and necessitates the inclusion of higher-order parasitic terms in modeling multi-qubit gate operations as $\Gside$ enters the megahertz regime.

\bibliography{lattice_5q}

\end{document}